\documentclass[journal]{IEEEtran}
\usepackage[linesnumbered,vlined,ruled,commentsnumbered]{algorithm2e}
\usepackage{algpseudocode}
\makeatletter
\def\therule{\makebox[\algorithmicindent][l]{\hspace*{.5em}\vrule height .75\baselineskip depth .25\baselineskip}}%

\newtoks\therules% Contains rules
\therules={}% Start with empty token list
\def\appendto#1#2{\expandafter#1\expandafter{\the#1#2}}% Append to token list
\def\gobblefirst#1{% Remove (first) from token list
  #1\expandafter\expandafter\expandafter{\expandafter\@gobble\the#1}}%
\def\LState{\State\unskip\the\therules}% New line-state
\def\pushindent{\appendto\therules\therule}%
\def\popindent{\gobblefirst\therules}%
\def\printindent{\unskip\the\therules}%
\def\printandpush{\printindent\pushindent}%
\def\popandprint{\popindent\printindent}%

\algdef{SE}[WHILE]{While}{EndWhile}[1]
  {\printandpush\algorithmicwhile\ #1\ \algorithmicdo}
  {\popandprint\algorithmicend\ \algorithmicwhile}%
\algdef{SE}[FOR]{For}{EndFor}[1]
  {\printandpush\algorithmicfor\ #1\ \algorithmicdo}
  {\popandprint\algorithmicend\ \algorithmicfor}%
\algdef{S}[FOR]{ForAll}[1]
  {\printindent\algorithmicforall\ #1\ \algorithmicdo}%
\algdef{SE}[LOOP]{Loop}{EndLoop}
  {\printandpush\algorithmicloop}
  {\popandprint\algorithmicend\ \algorithmicloop}%
\algdef{SE}[REPEAT]{Repeat}{Until}
  {\printandpush\algorithmicrepeat}[1]
  {\popandprint\algorithmicuntil\ #1}%
\algdef{SE}[IF]{If}{EndIf}[1]
  {\printandpush\algorithmicif\ #1\ \algorithmicthen}
  {\popandprint\algorithmicend\ \algorithmicif}%
\algdef{C}[IF]{IF}{ElsIf}[1]
  {\popandprint\pushindent\algorithmicelse\ \algorithmicif\ #1\ \algorithmicthen}%
\algdef{Ce}[ELSE]{IF}{Else}{EndIf}
  {\popandprint\pushindent\algorithmicelse}%
\algdef{SE}[PROCEDURE]{Procedure}{EndProcedure}[2]
   {\printandpush\algorithmicprocedure\ \textproc{#1}\ifthenelse{\equal{#2}{}}{}{(#2)}}%
   {\popandprint\algorithmicend\ \algorithmicprocedure}%
\algdef{SE}[FUNCTION]{Function}{EndFunction}[2]
   {\printandpush\algorithmicfunction\ \textproc{#1}\ifthenelse{\equal{#2}{}}{}{(#2)}}%
   {\popandprint\algorithmicend\ \algorithmicfunction}%
\makeatother

\usepackage{pifont}
\usepackage{graphicx}
\usepackage{amsmath}
\usepackage{ amssymb }
\usepackage{amsthm}
\usepackage[font=small,labelfont=bf,tableposition=top]{caption}
\usepackage{cuted}
\usepackage{subfig}
\RequirePackage{tikz}
\RequirePackage{zref-abspage}
\RequirePackage{zref-user}
\RequirePackage{tikz}
\RequirePackage{atbegshi}
\usetikzlibrary{calc}
\RequirePackage{tikzpagenodes}
\RequirePackage{etoolbox}
\RequirePackage{etoolbox}
\makeatletter
\newlength{\ALG@continueindent}
\newcommand*{\ALG@customparshape}{\parshape 2 \leftmargin \linewidth \dimexpr\ALG@tlm+\ALG@continueindent\relax \dimexpr\linewidth+\leftmargin-\ALG@tlm-\ALG@continueindent\relax}
\newcommand*{\ALG@customparshapex}{\parshape 1 \dimexpr\ALG@tlm+\ALG@continueindent\relax \dimexpr\linewidth+\leftmargin-\ALG@tlm-\ALG@continueindent\relax}
\apptocmd{\ALG@beginblock}{\ALG@customparshape\everypar{\ALG@customparshapex}}{}{\errmessage{failed to patch}}
\makeatother
\newtheorem{thm}{Theorem}

\newtheorem{cor}[thm]{Corollary}
\newcommand{\nosemic}{\renewcommand{\@endalgocfline}{\relax}}% Drop semi-colon ;
\newcommand{\dosemic}{\renewcommand{\@endalgocfline}{\algocf@endline}}% Reinstate semi-colon ;
\let\oldnl\nl% Store \nl in \oldnl
\newcommand{\nonl}{\renewcommand{\nl}{\let\nl\oldnl}}% Remove line number for one line
\makeatother
\begin{document}

%
% paper title
% can use linebreaks \\ within to get better formatting as desired
\title{Optimizing Capacitated Vehicle Scheduling with Time Windows: A Case Study of RMC Delivery }

% author names and affiliations
% use a multiple column layout for up to three different
% affiliations
\author{Mohamed~E.~Masoud,
        ~and~ Saeid~Belkasim
 % <-this % stops a space
\thanks{ Manuscript received... }
\thanks{..}
\thanks{M. Masoud, and S. Belkasim are with the Department
of Computer Science, Georgia State University, Atlanta,
GA, 30302 USA E-mail: mmasoud1@student.gsu.edu; sbelkasim@gsu.edu.}% <-this % stops a space
%\thanks{Reserved}% <-this % stops a space

}

%*********************

%********************
% conference papers do not typically use \thanks and this command
% is locked out in conference mode. If really needed, such as for
% the acknowledgment of grants, issue a \IEEEoverridecommandlockouts
% after \documentclass

% use for special paper notices
%\IEEEspecialpapernotice{(Invited Paper)}

% make the title area
\maketitle

\begin{abstract}
%\boldmath
Ready Mixed Concrete Delivery Problem (RMCDP) is a multi-objective multi-constraint dynamic combinatorial optimization problem. From the operational research prospective, it is a real life logistic problem that is hard to be solved with large instances. In RMCDP, there is a need to optimize the Ready Mixed Concrete ( RMC)  delivery by predetermining  an optimal schedule for the  sites-trips assignments that  adheres to strict time, distance, and capacity constraints. This optimization process is subjected to a domain of objectives ranging from achieving maximum revenue to minimizing the operational cost. In this paper, we analyze the problem based on realistic assumptions and introduce its theoretical foundation. We derive a complete projection of the problem in  graph theory, and prove its NP-Completeness in the complexity theory, which constitutes the base of the proposed approaches. The first approach is a graph-based greedy algorithm that deploys dynamic graph weights and has polynomial time complexity. The second approach is a heuristic-based algorithm coupled with the dynamic programming and is referred to as Priority Algorithm.  This  algorithm is carefully designed to address the RMCDP dynamic characteristic, and satisfies its multi-objectivity. In comparison with the state-of-arts approaches, our algorithm achieves high feasibility rate, lower design complexity, and significantly lower computational time to find optimal or very slightly suboptimal solutions.
\end{abstract}
\IEEEpeerreviewmaketitle


\section{Introduction}
% no \IEEEPARstart
The importance of the Ready Mixed Concrete Delivery Problem (RMCDP) stems from the huge construction industry that is based on the RMC production and dispatching processes. In US, the  RMC revenue in 2014 has been estimated as $\$30$ billion from dispatching around 301 millions cubic yards produced by around 5,500 ready mixed Concrete Batch Plants (CBP) and delivered by approximately 65,292 trucks as reported by the National Ready Mixed Concrete Association (NRMCA) in their annual fleet benchmarking survey [1]. The NRMCA report covers the operational data of 90 RMC companies from eight US geographical regions. This realistic RMC statistical data is the basis for our assumptions. 
In the RMCDP with one depot,which is our case of study, there is exactly one CBP that mixes RMC ingredients (e.g. cement,water,and aggregates) before loading the RMC products to a number of trucks in order to dispatch and haul the products to a number of construction sites. Each site must be  accessible by the depot within a specific time because of the perishable nature of the RMC product. This nature   stipulates that the consecutive deliveries per same site must not exceed a predefined time constraint in order to guarantee proper bonding between these consecutive deliveries and to avoid the occurrence of  planes of weakness in concrete or the so-called cold joints.  Therefore, this time constraint is important to prevent the concrete from reaching its initial setting and generates those joints. This initial setting time is defined by the standards specification for the RMC [2] and is considered as the upper bound of the RMCDP time constraint. Therefore, the RMC delivery in realistic is a hard scheduling problem for the multiple constraints it has. Among these constraints, there are the truck capacity constraint due to the limited drum size, the CBP capacity constraint due to the limited CBP mixer size, the travelling distance constraint from the depot to the sites due to the perishable nature of the product, besides the limited loading time slots for the trucks at the depot which is another constraint due to the finite number of these loading time slots, finally, there is the main time constraint that exists between the consecutive deliveries per site. Therefore, solving the RMCDP by finding an  effective dispatching algorithm is important for  optimizing the dispatching process by finding an effective feasible solution that can handle the mentioned constraints and meets the firm objective in a competitive way.

 %1-important question, is RMCDP routing problem or scheduling problem?, important note: there are three classes of the problems: P,NP,NPC. any problem can be solved in time O(n^k)for some constant k, where n is the size of the input to the problem, is class P probem which means it is solvable in polynomial time. %
The classification of the RMCDP has multiple points of view and whether or not it can be classified as a special case of the Vehicle Routing and Scheduling Problem (VRSP). The main routing characteristics of RMCDP that differentiates it from other VRSP subclasses, is the site-trip relation which is one-to-many in the RMCDP. This difference is due to the production capacity and the truck capacity limitations in the RMCDP. Another important difference in the RMCDP  is the limited number of the sites to be serviced, but this decrease in the total number of sites is usually coupled with an increase in the number of trips required for each site to satisfy its demand which make the problem hard to be solved. The problem feasible solution must satisfy the mentioned capacities and time constraints. 

In this paper, we go beyond the scope in [3] and  address the RMC delivery problem from different aspects in order to mine its main characterises and propose the proper solution approaches based on our analysis and theoretical foundation. Our main contributions are summarized as follows:

\begin{itemize} 
\itemsep0em 
 
 \item[$\bullet$ ] We provide a complete projection of the problem in  graph theory.
  \item[$\bullet$ ] We design  heuristic algorithm based on  priority principle of design.
   
  \item[$\bullet$ ] The test results show that our approach is competitive and able to find optimal or slightly suboptimal solutions with better processing speed and lower design complexity.
\end{itemize}

         The rest of the paper is organized as follows. In Section II, related work review is given. The problem analysis, definitions, modelling, graph representation, complexity theory reduction, and algorithm design are discussed in Section III. The implementation and results are shown in Section IV, and the conclusion is presented in Section V.
         
\section{RELATED WORK} 
     
       The VRSP is the main category in which the RMCDP can be sub-classified. VRSP has received an intensive research work for decades for the importance it has in advancing the logistics and fleet management processes. In addition to these areas, also supply chain management systems and just-in-time production strategies have been enhanced by the improvement in combinatorial-based routing and scheduling problems. Such improvements have showed a direct economic impact in all these fields and their related systems [4].

     Our consideration of the Vehicle routing problem (VRP) category and its subclasses (e.g. VRP with time window(VRPTW)) is because we show in this paper that the single depot RMCDP, which is our case of study, can be reduced in the complexity theory to that category of routing problems. Therefore, this reduction can mutually be exchanged between both domains  and the advances in any can possibly be propagated to the other. In VRP, many approaches have been used to find an optimal or near optimal solutions for the vehicle routing and scheduling problems, and some exact methods have been proposed [5]. Despite the preference of the exact solutions, they often perform poorly in the average and large solution spaces as shown by Kritikos et al [6]. Therefore, near-optimal approaches have been widely used as successful alternatives for their reasonable feasibility rate and low time complexity [7]. In  near-optimal cases, heuristic and meta-heuristic techniques constitute the main algorithm design solutions. Both have been used for solving the VRP in general and RMCDP in particular. In RMCDP, the evolutionary algorithms  have been widely adopted   [8-13]. For example, Feng et al [10] used Genetic Algorithm (GA) to solve the problem, while Cao and Lu [11] combined the genetic algorithm with simulation for more optimization. Particle Swarm Optimization (PSO) was used by Pan et al [12], and the Bee Colony Optimization (BCO) by Srichandum [13]. Despite the different source of inspiration for each technique, they have a high level of similarity in their basic ideas and share almost the same level of design complexity.
      
      The drawback of the evolutionary algorithms in the  RMCDP is that they try to select a number of random permutations  from the RMCDP permutation-based solution space when this solution space is actually distributed randomly for all the objective functions. Moreover, the correlation between the permutations in such a solution space is weak, which means that  a near neighbour of the worst permutation according to the objective function can be the optimal solution. Therefore, techniques such as mutation or crossover may not be considered  optimal in such space.     
      
       The other approach that is widely  used also in this domain is the linear programming based approaches [14-15]. In Yan and Lai [15], they deployed a mixed integer network flow model to solve the RMC delivery problem. Despite the feasibility of this approach, it brings additional complexity to the problem in terms of the large number of parameters used and the high computational times that are needed in case of sub-average problem instances.
       
         % { The current trend in CPU architectures is that improvements in clock frequencies are beginning to stagnate and chip makers are placing multiple cores in their CPUs in order to improve performance. The top-level workstation CPUs from AMD and Intel today have two cores on the CPU and CPUs with even 4 or even 8 cores are on Intel’s road map. In a few years single core CPUs might become obsolete. In order to get the full performance from these multi-core CPUs one needs to consider parallel programming. It is going to be interesting to see how big an impact this development is going to have on the heuristic community }

\section{RMC DELIVERY PROBLEM}
The main processes in RMC delivery operation are given in Fig.1. The  flow of the RMC delivery is represented by cyclic trips start and end at the depot during the RMC deliveries. In this paper, RMCDP refers to the problem that has exactly one depot for product loading,   and object $depot$ refers to the RMC factory that has exactly one $CBP$, or simply one $plant$, at which the mixing of the RMC ingredients and the loading of the RMC product to the trucks take place. The depot also has, beside the plant, the fleet for the RMC delivery and the other supporting systems (e.g. workshops, auxiliary equipment). Therefore,the object $depot$ is more wider than the object $plant$ in this context.

\begin{figure}[!h]
\centering 
  \includegraphics[scale=0.53]{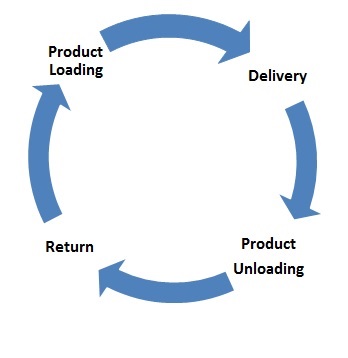}
  \caption{Depot-Site delivery flow digram starts by the product loading at the depot and represents the main components of the object $trip$.  }
  \label{fig_RMC}

\end{figure}
At this point also, we define the term $trip$ as the total time which includes : the truck loading time at the depot, the truck hauling time to the job site, the unloading process time at the placement point, and finally the returning time to the depot. Therefore, the $trip$ as a concept is the same for all sites, but as a parameter, it depends on the site distance from the depot and some other factors will be discussed later.
To avoid possible confusion and maintain consistency with the RMCDP domain vocabularies, the word $truck$ replaces vehicle, and $site$ replaces job-site from now on, and in this context,  object $site$ is the site that has exactly one RMC placement point.  [1-2].

% **********************************************
\begin{table}[t]
\caption{List of variables and parameters.  } % title of Table
\centering % used for centering table

\begin{tabular}{c c l c   } % centered columns (2 columns)
\hline\hline \\ %inserts double horizontal lines
Symbol  && Description  \\[0.5ex] % inserts table
%heading
\hline\\  % inserts single horizontal line

$n$ && Number of sites to be serviced\\

$k_{i}$&& Sequence of trips per site $i$\\

$k^{t}_{i}$&& Trip duration time per site $i$\\

$k^{s}_{ij}$&& Starting time at site $i$ for trip $j$.\\ 
$k^{s}_{i}$ &&Site $i$ proposed time by customer for first trip  \\
$k^{s}_{i1}$ &&Site $i$ first trip time at the site \\
$k^{d}_{ij}$&& Starting time at depot for trip $j$ of site $i$\\ 

$k^{e}_{ij}$&& Ending time at depot for trip $j$ of site $i$\\

$q_{i}$&& Demand of site $i$\\

$Q$&& Actual truck capacity for the homogeneous fleet\\

$K$&& Sequence of trips for all sites\\

$C$&& Depot concrete batch plant actual capacity\\
$D^{s}$&& Depot starting time\\

$P_{r}$&& Plant productivity\\
 
$L_{t}$&& Truck Loading time at the depot\\

$d_{i}$&& Site $i$ distance from the depot\\

$v_{i}$&& Truck average speed to Site $i$\\
 
$U_{i}$&& Truck unloading time at site $i$\\

$\gamma$&& Time of the initial setting of RMC = 90 min\\

$m$&& Total number of trucks per depot\\

$m_{u}$&& Upper bound of the trucks number needed for a delivery\\

$m_{l}$&& Lower bound of the trucks number needed for a delivery\\

$K_{s}$&& Total solution space of problem instance $I$\\

$K_{0}$&& Initial sequence of trips in  RMCDP graph\\

$G_{|V|}$&& Complete RMCDP graph\\

$L\left(v \right) $&&  Mapping each vertex $v$ to an element in $K_{0}$\\
 
$c_{v}$&& Cost of vertex $v$ in RMCDP graph\\

$c^{e}_{uv}$&& Cost of the edges between $\lbrace u,v\rbrace$ in RMCDP graph\\

$v^{s}$&& Service starting time at vertex $v$\\
$h_{L\left(v \right) }$ && Hauling time of the  site that is labelled to vertex $v$\\
$U_{L\left(v \right) }$ && Unloading time of the site that is labelled to vertex $v$\\
$s^{s}_{L\left(v \right) }$ && Proposed starting time of the site that is labelled to $v$\\

$c_{r}$&& Total cost of Humiliation circuit $r$ in RMCDP graph\\\\
    
\hline %inserts single line 
\end{tabular}
\begin{flushleft}
\begin{center}
\end{center}
\end{flushleft}
\label{table:nonlin} % is used to refer this table in the text
\end{table}
%--------------------------------
%Because of the absence of standardization in the notations and definitions of the problem, we define the RMCDP with a single depot and homogeneous trucks as follows:
Based on the previous basic description, the main $ characteristics$ that distinguishes the RMCDP from other vehicle scheduling and routing problems with time window constraint can be defined as follows:
\\\\
\textbf{Definition 1}. In single depot RMCDP with homogeneous trucks, we are given a single depot that has exactly one plant of capacity $C$, given a set of $n$ sites such that $n\geq2$, each site $i \in\lbrace 1,..,n\rbrace $ has an accessible distance $d_{i}$ from the depot and has a positive demand $q_{i}$. This demand to be satisfied needs  a sequence of trips $k_{i}=\left(  k_{i_{1}},..,k_{i|k_{i}|}\right)  $, $|k_{i}|\geq1$ from the depot by using  a set of $m$ homogeneous trucks each of capacity $Q$ and average speed $\overline{v}_{i}$ such that $Q < q_{i}$. 
Each trip is assigned to exactly one site by using exactly one truck, and the time lagging between
any consecutive trips $k_{i{j}},k_{i{j+1}}$ for site $i$ must not exceed the product initial setting time $\gamma$ such that the trips starting time at site $k^{s}_{i{j+1}}-k^{s}_{i{j}}\leq \gamma$.
The task is to find the best legal sequence of trips for all sites $K$ that can optimize the problem for the objective of minimizing the sites idling time awaiting for their deliveries while avoiding queue of trucks at the sites.

%In RMCDP, we are given a set of $n$ sites, such that $n\geq2$, each site $i \in\lbrace 1,..,n\rbrace $ has a positive demand $q_{i}$ and needs a sequence of trips $k_{i}=\left(  k_{i_{1}},..,k_{i|k_{i}|}\right)  $ from the depot to satisfy that demand,  such that the start time of the trip $k^{s}_{ij}<k^{s}_{ij+1}$ , $|k_{i}|\geq1$ and $k_{i}\subseteq K$, given a set of $m$ trucks each of capacity $Q <q_{i}$, each trip needs exactly one truck, and each truck visits exactly one site during that trip.

From definition 1, we can say that, as a general case in RMCDP, the truck capacity $Q <q_{i}$   $\forall i \in\lbrace 1,..,n\rbrace $. Therefore, for each site $i$, the site demand $q_{i}$ is partitioned into a set of $|k_{i}|$ elements such that $q^{s}_{i}=\lbrace q^{s}_{i_{1}},..,q^{s}_{i|k_{i}|}\rbrace $, where $q^{s}_{ij}\leq Q$ $ \forall j \in\lbrace 1,..,|k_{i}|\rbrace $. $|k_{i}|$ is the total number of trips for site $i$ that can be calculated as follows:
\\
\begin{equation}
|k_{i}|=\lceil\frac{q_{i}}{Q}\rceil   \quad\quad\forall i \in\lbrace 1,..,n\rbrace
\end{equation}
%The single depot RMCDP is defined as follows: Given a depot and a set of destinations with fixed distances from the depot, and given the average travel time per each destination, given %
Based on (1), we can formulate the total number of trips  $|K|$ for all sites as follows:
\\
\begin{equation}
|K|=\sum\limits_{i=1}^n |k_{i}|
\end{equation}
\\
In our analysis, we consider only the case of $homogeneous$  fleet in which all trucks have the same capacity $Q=\overline{Q} $. We mean by the truck capacity, the actual maximum capacity percentage of the truck gross drum volume according to the RMC standard [2]. Our assumption of the trucks homogeneity is mainly assumed to avoid adding another optimization problem which is the subset sum problem in the heterogeneity case,   also, truck homogeneity with maximum capacity is a desired goal for some economic factors that are related to truck  maintenance and operation. The ceil delimiter in (1) is used because in some cases the site demand $q_{i}$ or the RMC quantity of the site last trip $k_{i|k_{i}|}$  can be less than the truck capacity $Q$. However, in order to maintain the QoS in the RMC delivery operation, the loading time slot at the depot can exactly be assigned to one trip with one truck for one site delivery even if a site that has a total or partial demand less than $Q$.

%\textbf{Definition 2}. In single depot RMCDP with homogeneous trucks, we are given a single depot that has exactly one plant of capacity $C$, given a set of $n$ sites such that $n\geq2$, each site $i \in\lbrace 1,..,n\rbrace $ has an accessible distance $d_{i}$ from the depot and has a positive demand $q_{i}$. This demand needs to be satisfied by a sequence of trips $k_{i}$ using  a set of $m$ homogeneous trucks each of capacity $Q$ and average speed $\overline{v}_{i}$ such that $Q < q_{i}$. Each trip is assigned to exactly one site by using exactly one truck, and the time lagging between any consecutive trips $k_{i{j}},k_{i{j+1}}$ for site $i$ must not exceed the product initial setting time $\gamma$ such that the trips starting time $k^{s}_{i{j}}-k^{s}_{i{j+1}}\leq \gamma$. The task is to find the best legal sequence of trips $K=\left(  K_{1},..,K_{|K|}\right)  $ for all sites that can optimize the problem for the objective of minimizing the sites idling time awaiting for their deliveries while avoiding queue of trucks at the sites.

In definition 1, the $plant$ capacity $C$ represents the  volume of the plant mixer used for mixing the product ingredients. In general, for plant capacity $C < Q$, loading a truck with a trip quantity  $q^{s}_{ij}>C$ needs a set of mixer batches $B=\lbrace b_{1},..,b_{|B|}\rbrace$ such that $b_{i} \leq C$. For a realistic assumption, we assume that the plant capacity $C$ is the actual capacity used for the concrete batches during the truck loading process.

Under this assumption, and using the actual capacity parameter $C$, the plant productivity per hour can be determined by $P_{r}$  $ \leq C*60$  $m^{3}h^{-1}$ . In some cases,there are other factors that may impact the plant productivity such as increasing the batch mixing time at the plant mixer. This increase in  mixing time is important in some cases for improving the product quality [16]. However, above the $default$ mixing time value predefined by the mixer manufacture,  the  increase in mixing time results in decrease in the plant productivity according to its $nomogram $ [17]. One possible solution to maintain the same product quality without affecting the production rate is by distributing the mixing time between the plant and the truck mixers which is well-known as shrink-mixing  [2]. Based on these facts, we can neglect this factor, and consider only the actual capacity parameter $C$ for the $plant$ productivity.

Another variable in our model that also depends on the plant productivity $P_{r}$ is the truck $loading$ time $L_{t}$, which is proportional to the truck capacity $Q$ and inversely proportional to the plant productivity $P_{r}$ as follows:
\\
\begin{equation}
 L_{t}= \frac{Q}{P_{r}}*60
\end{equation}
The truck loading time variable $L_{t}$ is needed for two objects: the $trip$ object, and the $truck$ object. For the $trip$ object, each trip $k_{ij}$ has a $starting$ time $k^{d}_{ij}$, $end$ time $k^{e}_{ij}$, and trip $duration$ time $k^{t}_{i}$ for each site $i$ such that
 $k^{t}_{i}=\lbrace k^{e}_{ij}-k^{d}_{ij}|i\in\lbrace 1,..,n\rbrace,j\in\lbrace 1,..,|k_{i}|\rbrace\rbrace$, and based on our initial definition above for the object $trip$, we can formulate the trip duration $k^{t}_{i}$ as follows: 
 \begin{equation}
 k^{t}_{i}= L_{t}+ 2\left( \frac{d_{i}}{\overline{v}_{i}}\right)+ U_{i}
\end{equation}
\begin{equation}
 k^{e}_{ij}= k^{d}_{ij}+k^{t}_{i}
\end{equation}
\\
For simplicity, we assume in (4) that both of the hauling time and returning time are close to each other, that is, their difference can be neglected. Also, the other minor tasks that are associated with the RMC delivery are implicitly embedded in (4). These minor tasks may include the lab tests (e.g. slump test) and the time needed for rinsing the truck drum  after the unloading the product in the site. The variable $U_{i}$ in (4) is the unloading time at site $i$. 

 Equation (5) represents the sequence of trips for each site $i$ where site $i$ first trip starting time   $k^{s}_{i_{1}}$ should match the site proposed starting time $k^{s}_{i}$, where $k^{s}_{i_{1}}=k^{d}_{i{1}}+L_{t}+\frac{d_{i}}{\overline{v}_{i}}$.

For the $truck$ object, the need for the parameter $L_{t}$  is to determine the $upper$ bound of the number of trucks $m_{u}$ that should be $available$ for the RMC deliveries as follows:
\\
\begin{equation}
 m_{u}= \frac{2.\gamma}{L_{t}}
\end{equation}
\\
The time window $\gamma$ in definition 1 and (6) represents the maximum  time allowed between any two consecutive trips for the same site in order to maintain the product workability before the RMC reaches its initial setting. This time is estimated as 90 minutes under normal working conditions according to the RMC standard specifications [2].  

We assumed also the truck speed $\overline{v}_{i}$ depends on the site $i$ location for the fact that some sites can be in high density areas and others may not. Usually there are pre-delivery arrangements between the depot and the new site such as determining the best route to the site, and the average truck speed to estimate the product hauling time to the site. 

\subsection{Problem Classification}

Based on previous analysis and modelling for the RMC delivery problem, we can state the following: 
\\\\
\textbf{Definition 2.} The RMCDP is a multi-constraint and  multi-objective  optimization problem.

From definition 1  we can identify  different types of constraints in the problem. Starting with the time, the RMC time window $\gamma$ enforces a time constraint between the consecutive trips of same site $i$ such that $\left( k^{s}_{ij+1}-k^{s}_{ij}\right)\leq \gamma $. Also a lower bound for the time lag between the same trips can be generated as an additional constraint if, for example, the objective of the optimization problem is to minimize the trucks waiting time at site $i$ such that $U_{i}\leq\left( k^{s}_{ij+1}- k^{s}_{ij}\right)\leq \gamma $.  Another constraint can also be generated by $\gamma$ is the radius of the depot service area such that $L_{t}+\left( \frac{d_{i}}{\overline{v}_{i}}\right)+U_{i} \leq \gamma$ , where $ \left( \frac{d_{i}}{\overline{v}_{i}}\right) $ represents the RMC hauling time to   site $i$ for each $i\in\lbrace1,..n\rbrace$. Other time constraints can also be generated based on the problem objective. Beside the time constraints, there is also a capacity constraint that gives the RMCDP its unique characteristics and also its complexity. The limited truck capacity $Q$ compared to the site demand $q_{i}$   creates a constraint on the maximum number of visited sites per trip and limits that number to be exactly one site visit by one truck per trip. Other capacity  constraints can also exist depending on the objective function, for example, if the objective is to maximize the number of sites to be serviced in specific time unit, this will result in creating a constraint on the maximum number of trips per site at that time unit. Many other examples also can be given here as a proof of the multi-constraint and multi-objectivity of the problem.  

 %regardless of the objective of the problem, the solution space is the same in every case because it only depends on the number of sites $n$ and the number required trips per site $|k_{i}|$%
 
It is easy to prove that trying to solve a problem instance for different objectives results in the same solution space but not the same solution. Also, running the same optimization problem under different number of constraints affects only the number of feasible solutions not the size of  solution space. An example of that is the RMCDP in which its solution space size was originally estimated by Feng et al [10] and modified here to suit our assumptions and notations.
In RMCDP, the solution space of all possible permutations of a problem instance $I$ can be represented by the set $K_{s} =\lbrace K_{s_{1}},..,K_{s_{|k_{s}|}} \rbrace$ such that the best solution sequence $K\subset K_{s}$. Therefore, for any instance $I$, the size of its solution space  $|K_{s}|$ can be stated as follows:  
\\
\begin{equation}
|K_{s}|=\frac{(\sum\limits_{i=1}^n |k_i|)!}{ \prod\limits_{i=1}^n (|k_i|!)}
\end{equation}
\\
From (7), we can say that the total number of permutations $ |K_{s}|$ is a function only in the number of sites $n$ and the number of trips per site $|k_{i}|$.
For illustration, a simple example which we refer to  as   $example-1$   is given as follows: suppose we are given two sites with two trips per site, then $|K_{s}|=\lbrace6\rbrace$ possible sequences of trips such that $K_{s} =\lbrace \left( 1,1,2,2\right) ,..,\left( 2,2,1,1\right)\rbrace$. One of these sequences should be the winner sequence $K$ that best meets the problem objective. Therefore, the preference of the best feasible solution in the RMCDP depends on the performance of the trips dispatching sequence according to the problem objective. The best sequence is nothing but a permutation from the solution space. Therefore, designing an efficient algorithm to find this optimal permutation is the main challenge that needs to be tackled. This challenge stems from the huge solution space that exists when many sites with many trips per site are proposed. For a realistic example, let us reconsider the RMC  statistical data reported by the NRMCA [1], in which there are  around 301 millions $yd^{3}$ of RMC produced by 5,500 plants in one year.  From this data we can calculate an average of 210 $yd^{3}$ of RMC as a daily production rate by each plant. By considering also, the average truck load capacity in the survey which is 8.0 $yd^{3}$, then we have around 26 trip from each plant per day. Such a number of trips can result from, for instance, five sites with five trips per site. These numbers of sites and their trips  can produce a total solution space of more than 600 trillions of  possible trips sequences. If we neglect the memory limitation, a processor with speed of 5 $GHz$, if exists, can find the optimal solution for this average size problem in around 32 hours, which in reality cannot be acceptable as a practical solution for the daily-based scheduling plan. Such a huge solution space and its high computational cost is expected in exact methods. Therefore, there is a need to identify the problem class precisely and which solution strategy should be adopted and why.

\subsection{RMCDP in Graph theory}
%In related literature, the attempts to considering the RMCDP as m-TSP by considering the graph nodes as the sites/customers is not the right projection of the problem in the graph theory, nor the right reduction according to the complexity theory.
In this section, we address a common issue in the related literature, which is the absence of the problem projection in graph theory, the step that is imperative for a proper reduction of the problem. In complexity theory, and in order to identify the complexity class of the RMCDP, we need  to find a polynomial time reduction algorithm such that $any$ instance of RMCDP can be transformed into an instance of a well-known classified problem. Therefore, defining the problem in graph theory comes first in order to accomplish that reduction.  Based on our notations and prior definitions, we can state the following:
\\\\
\textbf{Definition 3}: In RMCDP, and for a problem instance $I$ and objective $f$, the solution space is given by the set $K_{s}$ of all possible  sequences of trips such that any solution sequence $K\in K_{s}$. The set $K_{s}$ can be represented by a  $complete$ graph $G_{|V|} =\left( V,E\right) $ with a weight function $w:E\rightarrow R$, and a set of vertices $V=\lbrace s,v_{1},..,v_{|K|} \rbrace$ such that vertex $s$ is the depot, and $|K|$ is the total number of trips for all sites.  Each $v\in V$ has a cost $c_{v}\geq 0$ depending on the objective $f$ and has a starting time  $v^{s}$ to be serviced.  Each $v\in V\setminus \lbrace s\rbrace$ is visited exactly once and assigned to exactly one trip element $\lbrace a_{\kappa}\rbrace$ from any $initial$ sequence of trips $K_{0}=\left(  a_{1},..,a_{|K_{0}|}\right) $ and Labelled with that element such that $L\left(v \right)=\lbrace a| a_{\kappa}\in K_{0}, \kappa\in\lbrace1,..,|K_{0}|\rbrace\rbrace$.
$E$ is the set of edges such that each edge $\lbrace i,j\rbrace $ has a cost $c^{e}_{ij}$ and associated with a time that can be defined as a function in the loading time $ L_{t} $.% The total cost of the solution %A binary decision variable is defined as $x_{ij}=\lbrace0,1\rbrace$ is used for considering only the used edges in the solution by setting the variable to one according to the problem objective function:
%\begin{equation}
%min\sum_{i\in V}\sum_{j\in V \setminus\lbrace i\rbrace} w_{ij}\left(t \right) x_{ij}   
%\end{equation}
%\\
%where $t\in\lbrace t_{0},t_{0}+L ,..,t_{0}+|K|L \rbrace$, $t_{0}=min\left(  k^{s}_{i1}\right)$ and the number of trucks available $m\geq 2\gamma.L^{-1}$. The weight function $w\left(t \right) $can defined as:

%\[
% w(t)=
% \begin{cases}
% w_{0}, & t=t_{0} \\
% f(w), & t>t_{0}
% \end{cases}
%\]
In definition 3, we define RMCDP  as a multi-objective problem such that a problem objective function $f\in \textbf{\textit{F}}$ , where $\textbf{\textit{F}}$ is a domain of objectives applicable to the RMCDP. Before discussing our  problem objective in this paper , and how to represent it in the graph theory, there is a need at first to give some insights on the RMCDP graph main characteristics. % there we discuss this part and show how to formulate the objective function in the graph theory, we need first to give some insights on the RMCDP graph main characterises and prepare it for a complete functionality.  for the minimization objectives such as minimizing the trucks idling time at the sites, the problem can be formulated as follows:
%\begin{equation}
%$min\sum_{v\in V}\sum_{u\in V \setminus\lbrace v\rbrace} c_{v,u} x_{vu} $  
%\end{equation}
%\\
%]where $x_{ij}=\lbrace0,1\rbrace$ is a  binary decision variable used for considering only the used edges in the $feasible$ solution by setting the variable to one in this case. Therefore, the need to identify a problem instance solution in the RMCDP graph coupled with the need to verify each solution feasibility is vital at this point.
\begin{thm}
In RMCDP completed graph $G$, for a problem instance $I$, any solution $K \in K_{s}$  is a $simple$ cycle in the graph $G$.
\end{thm}

\textbf{Proof:}  From definition 3, RMCDP graph $G=\left(V,E \right) $ is a $complete$ graph such that each vertex $v\in V\setminus\lbrace s\rbrace$ represents a trip element in $K$ and $|V|=|K|+1$, where $|K|$ is the total number of trips of all sites can be found by (1) and (2). Each sequence of vertices starts and ends at vertex $s$ and each $v\in V\setminus\lbrace s\rbrace$ is visited exactly once, which results in generating a $simple$ cycle in $G$. Let $S_{n}$ be the total number of all possible $simple$ cycles in $G$ that  start and end at vertex $s$,  $S_{n}=\left( |K|\right) \left(|K|-1\right)\left(|K|-2\right)..\left( 2\right) \left( 1\right) = |K|!$ which is a greater number than the total solution space $|K_{s}|$ in (7).  $\quad\blacksquare$

For illustrating these concepts, let us reconsider again the simple $example-1$ in the previous section in which we are given two sites with two trips per sites. By (2), the total number of trips for all sites $|K|=4$, and the solution space is given by a set $K_{s}$ of all possible sequences of trips such that $|K_{s}|=6$ . According to our definition (3), we can formulate this simple problem instance in  graph theory using a complete graph $G_{|V|}$ such that $|V|=|K|+1=5$. The initial sequence $K_{0}$ can be any sequence of trips for all sites as $K_{0}=\left(1,1,2,2 \right) $.  $K_{0}$ can be represented by $G_{5}$ such that each vertex $v\in V\setminus\lbrace s\rbrace$ is labelled with a trip elements in $K_{0}$ as shown in Fig. 2. The figure shows six   $simple$ cycles in $G_{5}$ that represent the problem  instance solution space $K_{s}$. 
\begin{figure}[t]
\centering 
  \includegraphics[scale=0.45]{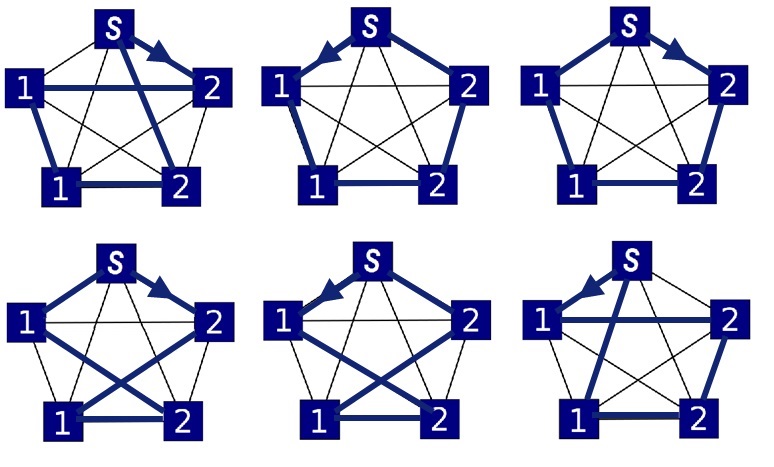}
  \caption{Example-1 solution space where all sequences of trips are represented by simple cycles in RMCDP graph $G_{5}$  }
  \label{fig_RMC}

\end{figure}
\\
\begin{cor}
A solution is $feasible$ in the RMCDP graph $G_{|V|}$, if it is a $simple$ $cycle$ satisfies the RMCDP constraints.
\end{cor}
\textbf{Proof:}  Let $n$ be the total number of sites , $|k_{i}|$ the total number of trips for site $i$ where $ i\in \lbrace 1,..,n\rbrace $, let the $initial$ sequence of trips $K_{0}$ is any sequence of all trips for all sites such that each trip $k_{ij}\in K_{0}$, and for each $v\in V\setminus \lbrace s\rbrace$ in the RMCDP complete  graph $G$, let the label of each vertex be as follows:  
\\    
\begin{equation}
 L\left(v \right)=\lbrace i| k_{ij}\in K_{0}, i\in\lbrace1,..,n\rbrace,j\in\lbrace1,..,|k_{i}|\rbrace\rbrace
\end{equation}
\\
Now for any $simple$ cycle $r =\left(s,v_{1},v_{2},..,v_{|K_{0}|} \right) $ in   graph $G$, if any two vertices $ v \in r\setminus\lbrace s\rbrace,u \in r\setminus\lbrace s,v\rbrace$ have the same label such that  $L(v)=L(u)$, and the intermediate vertex $\nu_{m}$ between them (if exists) has different label such that $L(\nu_{m})\neq L(v)$,  then $r$ is a $feasible$ solution in $G$ if and only if  the same label  vertices starting times $v^{s}$ and $u^{s}$ satisfy the follows:
\begin{equation}
\left( v^{s}-u^{s}\right)  \leq \gamma
\end{equation}
where $u^{s} <v^{s}$, and $\gamma$ is the maximum time window allowed between $v^{s}$ and $u^{s}$.

For the capacity constraint, it is considered in the RMCDP graph $G$ by the bijective  mapping and labelling of each vertex $ v \in r\setminus\lbrace s\rbrace$ to each trip element in the initial sequence of trips $K_{0}$. For the number of trips $|k_{i}|$ per site $i$ and the total number of trips for all sites $|K|=|K_{0}|$ both are determined based on the truck capacity $Q$. 
$\quad\blacksquare$
%*******************Fig 3 ******************
\\\\
 \begin{strip}
    \centering
        \includegraphics[scale=0.5]{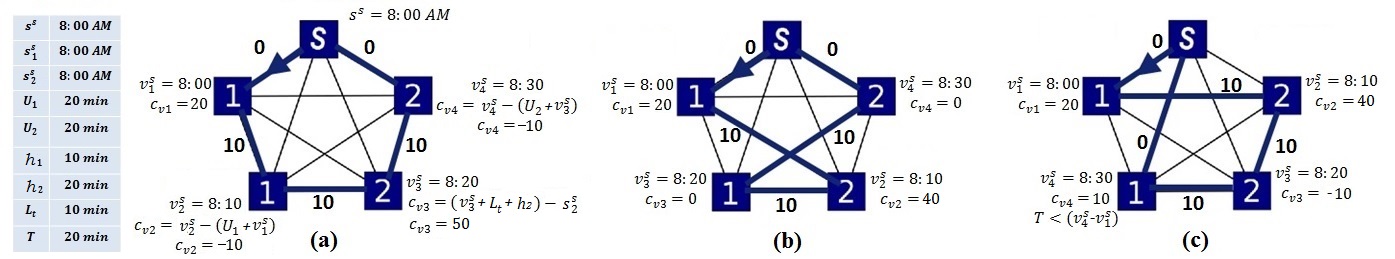}
        \captionof{figure}{Example-1 of three sequences of trips represented by simple cycles in RMCDP graph $G_{5}$. \textbf{(a)} The first solution sequence of trips $r_{1}$=(1,1,2,2) has a  total cost of sites waiting $c_{r_{1}}$=70 $min$, and a total truck idling time at the sites of 20 $min$. \textbf{(b)} The feasible solution sequence of trips $r_{2}$=(1,2,1,2) has a  total cost of   $c_{r_{2}}$=60 and zero truck idling time at the sites(\textit{best sequence}). \textbf{(c)} An \textit{infeasible} solution sequence of trips  (1,2,2,1) which results in a delay between the two consecutive trips for the same site (\textit{site 1}) such that $v^{s}_{4}-v^{s}_{1}>T$.  }  
        \vspace{-0.1cm}
  \end{strip}  
%***************************************** 
Under the assumption of the availability of enough number of trucks for all loading time slots at the depot, which we may refer to it as $ assumption-1$ , the service starting time $v^{s}$ for any vertices $v\in V\setminus\lbrace s\rbrace$ in the feasible solution $r$ depends on the depot starting time $s^{s}$ such that :
\begin{equation}
\begin{split}
 v^{s}_{i}&= s^{s}+\left( i-1\right)L_{t}
 \end{split}
\end{equation}
\begin{equation}
\begin{split}
 =s^{s}+c^{e}_{1i} 
 \end{split}
\end{equation}  
where $ c^{e}_{1i}$ by definition (3) is the cost of all edges between the vertices $v_{1}$ and $v_{i}$ in the simple cycle sequence $r$ such that:
 \begin{equation}
    c^{e}_{1i}=\sum_{j=1}^{i-1} c^{e}_{jj+1} 
 \end{equation}  
%where $i$ is the vertex order, $i\in \lbrace 2,..,|K_{0}|\rbrace$ in the feasible solution sequence $r$. Therefore, under assumption (1), the edge weight or cost  $c^{e}_{jj+1}$ in the RMCDP graph $G_{|V|}$ represents the truck loading time $L_{t}$ such that $c^{e}_{jj+1}= L_{t}$. Also if we consider the objective of minimizing the total trucks idling time at the sites, then we can formulate the cost $c_{v}\in\mathbb{R^{+}}$ of each vertex  in RMCDP graph $G_{|V|}$ as a function at the site unloading time $U_{i}$  as follows:
where $i$ is the vertex order, $i\in \lbrace 2,..,|K_{0}|\rbrace$ in the feasible solution sequence $r$. Therefore, under assumption (1), the edge weight or cost  $c^{e}_{jj+1}$ in the RMCDP graph $G_{|V|}$ represents the truck loading time $L_{t}$ such that $c^{e}_{jj+1}= L_{t}$. Also if we consider the objective of minimizing the total sites waiting time with no truck queues at sites, then we can formulate the cost $c_{v}\in\mathbb{R^{+}}$ of each vertex  in RMCDP graph $G_{|V|}$ as  follows:    
 \begin{equation}
c_{v}=
\begin{cases}
  
 v^{s}-\left(u^{s}+U_{L\left(v \right) }\right):&L\left( v\right) = L\left( u\right) \\
  \left(v^{s}+L_{t}+h_{L\left(v \right) } \right)-s^{s}_{L\left(v \right) }: & \left( u^{s}=s^{s}\right) \\
 0 & \left( c_{v}<0\right) 
\end{cases}
\end{equation}
If $ U_{L\left(v \right) }\leq\left(v^{s}-u^{s}\right)$, then there will be no truck queues at sites. $s^{s}_{L\left(v \right) }$ is the suggested starting time for the site that is labelled to vertex $v$, $h_{L\left(v \right) }$ is the hauling time of the site that is labelled to vertex $v$.  For any  intermediate vertex $\nu_{m}$ between $v$ and $u$ (if exists), it should have a different label such that $L(\nu_{m})\neq L(v)$. In other words, $v$ is the next similar label vertex to $u$ in the feasible solution $r$ and both represent two consecutive trips for the same site.

 In (13), if $\left( c_{v}<0\right)$ this means there is no site waiting time for the trip that is mapped to current $v$, but there is a truck idling time at the site for that trip, and this idling time duration is the same as the calculated $c_{v}$ of vertex $v$.
From (10) and (13) we can state the total cost $c_{r}$ for some feasible solution $r$ as follows:
 \begin{equation}
c_{r}= \sum_{i=1}^{|r|-1} c_{vi}  
\end{equation}
%\begin{equation}
%min\sum_{r=1}^{|K_{s}|} c_{r} x_{r}   
%\end{equation}
%where $x_{r}=\lbrace0,1\rbrace$ is a  binary decision variable used for considering only the used edges in the $feasible$ solution by setting the variable to one in this case. 
The best solution of the problem is the one that has minimum $c_{r}$ and satisfying its constraints.
For illustrating our findings, suppose for example-1 that the trips unloading time in minutes $U_{1}$ and $U_{2}$ for sites 1 and 2 is 20 for each, and the truck loading time $L_{t}$ at the depot is 10 under the assumption (1) stated above. Let the suggested starting time $s^{s}_{1}$ and $s^{s}_{2}$ for both sites be 8:00 AM, and the hauling time $h_{1}$ and  $h_{2}$ are 10 and 20 $min$. Suppose the maximum time allowed  $T$ between consecutive trips is 20 $min$ which is small value used for the  illustration purpose. With these parameters example-(1) graph $G_{5}$ can be represented by a weighted complete graph such that for each edge $e\in E\setminus \lbrace s,v\rbrace$   the cost $c^{e}=L_{t}$ and $\lbrace 0\rbrace$ otherwise, and for each vertex $v\in V\setminus\lbrace s\rbrace$, the cost $c_{v}$ determined by (13). The resulted graphs of three different solutions are shown in Fig. 3. The best feasible solution among the represented solutions is $\left(b \right)$ which has zero truck idling time at the sites and a total sites waiting time $c_{r_{b}}=60$. Because we address the objective of minimizing the sites waiting time with no truck queues or zero truck idling time at the sites, then  solution $\left(a \right)$ is infeasible because it results in a total trucks idling time of 20 $min$ from the sum of $c_{v2}$ and $c_{v4}$ where the cost of both are in negative. When the cost of a trip is in negative   $\left( c_{v}<0\right)$ as in (c), this means the truck of this trip is in the site waiting till the previous trip finish its unloading phase. Therefore, if a trip has a truck idling time at a site, then the cost of site waiting for this trip is zero as in (13).  For graph $\left(c \right)$, the solution sequence is infeasible for not satisfying the time constraint $T$. By the end of the previous example, a proper projection of the RMCDP in graph theory have been achieved, which is imperative step for the problem classification in complexity theory.
\begin{thm}
RMCDP completed graph $G_{|V|}$ is a $Hamiltonian$ graph, that is, RMCDP as a decision problem is NP-complete.
\end{thm}
\textbf{Proof:} In the RMCDP completed graph $G_{|V|}$ there are $\left(|V|-1 \right)!$ $Hamiltonian$ $cycle$ (HC) which we referred to them before as simple cycles. Therefore, any HC in the RMCDP graph  represents one possible solution sequence for the problem.
Now to prove the $NP-completeness$ of the RMCDP, we need first to put it in a $decision$ problem form such that for some problem instance $\langle G_{|V|},c_{r},W\rangle$ RMCDP can be defined as follows: Given a RMCDP instance $\langle G_{|V|},c_{r},W\rangle$ , and positive integer $W$, is there a feasible solution sequence such that its cost $c_{r}\leq W$ ? 
%*****************Fig 4 **************
\\\\\\
\begin{strip}
    \centering
        \includegraphics[scale=0.5]{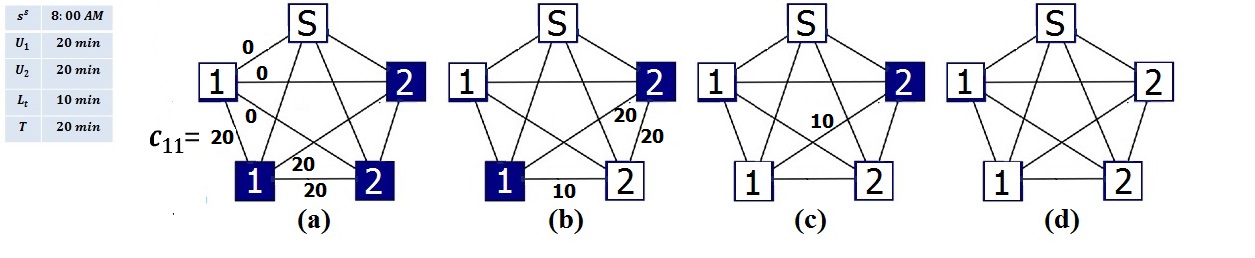}
        \captionof{figure}{Applying RMCDP Greedy Algorithm to Example-1 by initializing the edges costs to $\lbrace 0\rbrace$ as in (15).  \textbf{(a)} Starting from $s$ and vertex $L(v)=1$ by moving them to $V_{H}$, $c^{e}_{vu}=U_{u}=20$ when $L(u)=L(v)=1$. (16) is used to duplicate this cost to all the edges of $L(u)=1$. \textbf{(b)} $L(u)=2$ is selected to move to $V_{H}$ for the minimum edge cost it has as shown in previous step. According to (15), $c^{e}_{vu}=U_{2}=20$ when $L(u)=L(v)=2$ and $c^{e}_{vu}= c^{e}_{vu}-L_{t}=10$ when $L(u)=1$ because $L(u)\in V_{H} $. \textbf{(c)} $L(u)=1$ is selected to move to $V_{H}$, $c^{e}_{vu}= c^{e}_{vu}-L_{t}=10$ when $L(u)=2$ because $L(u)\in V_{H} $. \textbf{(d)} $L(u)=2$ is selected to move to $V_{H}$, the solution sequence is $\left(1,2,1,2 \right)$.}
  \end{strip}  
%**************************************** 
The answer of the decision problem is yes/no. In case the answer is (yes), then we have a sequence of vertices that can be $verified$ in polynomial time such that the verification algorithm checks: 1) each vertex exists exactly once in the solution sequence, 2) the vertices total cost is at most $W$, 3) the lagging time between any same label and consecutive vertices is less than or equal $\gamma$. These checks can be performed in $\Omega\left(|V| \right) $, that is, RMCDP$\in$ NP.
It is well known that finding HC in a graph is NP-complete problem, and it can be said that HC problem (HCP) is as hard as finding a minimum cost HC in a completed undirected graph because in both cases the solution space of the graph is $O\left(|V|! \right)$. Moreover, Solving the RMCDP is nothing but solving the HCP because the best solution in RMCDP graph is actually a HC that best meets the objective function. Thus, we can state that HCP $\leq_{p}$ RMCDP if this reduction satisfies the following: First, for any number of sites $n$, finding the initial sequence of trips $K_{0}$ as an input instance for the transformation function is $O \left( n\right) $ based on (1) which is used to determine the number of trips for each site.

   Also the mapping function in (8) which is used to transform the RMCDP instance to the completed Hamiltonian graph by assigning each trip element in the instance initial sequence of trips $K_{0}$ to each vertex in the graph, this function is a polynomial time  transformation function with a running time of $O \left( |K_{0}|\right) $. 
    
    In summary, because of the polynomial reduction  HCP $\leq_{p}$ RMCDP which we just proved, then RMCDP is NP-hard, and because RMCDP $\in$ NP as we also proved, that is, RMCDP is NP-complete.$\quad\blacksquare$

The importance of theorem (3) stems from the fact that it shows that most probably there is no exact algorithm is capable of solving the RMCDP in polynomial time. Another important property of the problem can be noticed from Fig. 3 is  the $dynamic$ characteristic of RMCDP because of its time dependency. Therefore, our solution strategies are designed based on  these facts.

In graph theory, a $greedy$ approach can be adopted to solve the RMCDP completed graph $G_{|V|}$ after considering the dynamic property of the problem. Hence, the graph edge cost $c^{e}_{vu}$ should not be considered as static value but a non-static value that changes in discrete time and is needed to be determined as a critical part in the proposed solution. For applying this solution, let the objective function be minimizing the sites waiting time with minimum trucks idling time at the sites. The greedy approach for the RMCDP can be stated as follows: Let the RMCDP graph $G_{|V|}=\left(V,E,w,t \right)$ be a weighted completed graph and $H_{G}=\left(V_{H},A_{H},w_{H} \right)$ be  a $simple$ $directed$ graph represents the minimum cost Hamiltonian circuit of $G_{|V|}$ such that $|V_{H}|=|A_{H}|$, the greedy algorithm $RMCDP\left( G_{|V|},H_{G}\right) $ can be stated as follows:
%******************
\begin{algorithm}
\DontPrintSemicolon
\SetKwInOut{Input}{Input}\SetKwInOut{Output}{Output}
\SetKwFunction{KwFn}{}
\Input{$G_{|V|}=(V,E,w,t)$, $G_{|V|}$ is RMCDP graph.}
\Output{$H_{G}=(V_{H},A_{H},w_{H})$, $H_{G}$ is $min\left( HC\right) $ of $G_{|V|}$.}
\BlankLine
\nonl\textbf{RMCDP ($G_{|V|},H_{G}$)}\;
$N  \longleftarrow |V|$\;
$V_{H}  \longleftarrow \emptyset$\;
$A_{H} \longleftarrow \emptyset$\;
$V_{H} \longleftarrow s$ :$V_{H}=V_{H}\cup\lbrace s\rbrace, V=V-\lbrace s\rbrace $\;
Move any $v\in V$ to $V_{H}$ : $V_{H}=V_{H}\cup\lbrace v\rbrace, V=V-\lbrace v\rbrace $\;
$A_{H}=A_{H}\cup \lbrace s,v\rbrace$, $E=E- \lbrace s,v\rbrace$.\;
Let $V_{H}$ be $sequence$ : $ V_{H}=\left(  s,v_{1}\right) $\;
\For{$i = 1$ \KwTo $N-2$}{
Let $\lbrace v_{i},u\rbrace$ be an edge such that $v_{i}\in V_{H}$, $u\in V$\;
\For {$each$ $ Distinct$ $u\in V$}{
$Find$ $min\left( c^{e}_{v_{i}u}\right)  $
}

$V_{H}= V_{H}\cup \lbrace u\rbrace$.\;
$A_{H}=A_{H}\cup \lbrace v_{i},u\rbrace$.\;
$V=V-\lbrace u\rbrace$.\;
}
\If{ $IsFeasible$ $\left( H_{G}\right) $}{
\Return $H_{G}$
}
\Else {$Print\left(   No Feasible Solution\right). $} 
\caption{RMCDP Greedy Algorithm\label{IR}}
\end{algorithm}
%**************
%*******************
\\
where $N$ is the total number of vertices in RMCDP graph $G_{|V|}$ in which each vertex $u\in V\setminus\lbrace s\rbrace$ represents a trip element as shown before. In RMCDP algorithm (1), the RMCDP main graph $G_{|V|}$ is converted to a simple directed graph $H_{G}$ that starts from the vertex $s$ and then collects the other trip elements vertices one by one in a greedy approach by considering the next $distinct$ vertex $u\in V$ from $G_{|V|}$ that has minimum cost with the last moved vertex $v_{i}$ in $H_{G}$ where  $c^{e}_{vu}$ is the edge cost between vertices $v$ and $u$ such that:

\begin{equation}
 c^{e}_{vu}=
 \begin{cases}
  0 & |V_{H}|\leq 1\\
  U_{u} & L(v)=L(u) \\
  c^{e}_{vu}- L_{t} & L(u)\in V_{H}
 \end{cases}
\end{equation}

\begin{equation}
 c^{e}_{uw}=\lbrace c^{e}_{vu}|  \forall w\in V\rbrace
 \end{equation}
\\
where $U_{u}$ is the unloading time of the next trip vertex $u\in V$, and $ V_{H}\setminus\lbrace s\rbrace$ is the visited vertices represents the trips that started when  $ |V_{H}|>1$ . At first, all the edges in $G_{|V|}$ have the same priority, and visiting any vertex results in updating $c^{e}_{vu}$ for each $u\in V$ according to (15). The last line in (15) stipulates visiting a previous $u$ with the same label of $v$ before updating $c^{e}_{vu}$. The iteration of the \textbf{for} loop in line 8 has a computational cost :  $\left( N-2 \right)+\left( N-2 -1 \right)+..+ 2 + 1 = \frac{\left( N-1 \right).\left( N-2 \right)}{2} $  , which is $ O\left ( N^{2} \right)$. This computational cost can be more optimized by considering only the $distinct$ vertices in each iteration that have different labels. In this case, such vertices can optimize the cost to $ O\left ( n^{2} \right)$, where $n$ is the total number of sites. Also, the feasibility verification function in line 15 has an average cost of $ O\left ( N^{2} \right)$ which results in a total time complexity for the RMCDP greedy algorithm of $ O\left ( N^{2} +n^{2} \right)\simeq O\left ( |K|^{2}\right)$  including the edges cost update after each iteration according to (15).

 Illustrating the performance of the RMCDP Greedy Algorithm is shown in Fig. 4. By applying the algorithm to Example-1, a feasible solution is reached with the minimum cost Hamiltonian cycle  and time complexity of O$\left ( |K|^{2}\right)$ where $|K|$ represents the total number of trips as in (2).  
%where $t\in\lbrace s^{s},t_{0}+L ,..,t_{0}+|K|L \rbrace$, $t_{0}=min\left(  k^{s}_{i1}\right)$ and the number of trucks available $m\geq 2\gamma.L^{-1}$. The weight function $w\left(t \right) $can defined as:
\subsection{RMCDP Priority Algorithm}
  
 Based on the previous analysis and the mathematical model for the RMCDP, this problem is a problem that belongs to  NP-complete (NPC) class, which means that the RMCDP is as hard as any problem in NP. All these problems that belong to NPC class are intractable and solving any of them results in solving the others. The advisable approach is to try for an approximation algorithm to solve NPC problem rather than searching for a polynomial time exact solution algorithm.  This option also has two challenges, the first one is the inefficiency of the designed algorithm, and the second, is the need to handle the dynamic property of the RMCDP. We classify it as a dynamic problem because of the time dependency it has. Therefore, our second strategy in designing the solution algorithm after the graph-based one is based on understanding the main characteristic of the problem that we use as a principle for the algorithm design.  
 
 \subsubsection{\textbf{Priority Algorithm - Site Waiting}}
 The problem objective that we use in last section and has been studied broadly in literature is the objective of minimizing the sites waiting times while maintaining the minimum trucks idling time at the sites, the situation that may occur if two consecutive trips or more for a site can cause a trucks queue during the product unloading at that site. The feasible sequence of trips for a given number of sites $n$ is the one that can satisfy the problem constraints as follows:
 \\
\begin{equation}
 U_{i}\leq\left( k^{s}_{ij+1}- k^{s}_{ij}\right)\leq \gamma     
\end{equation} 
\\
where the site $id$  $ i \in\lbrace 1,..,n\rbrace$, $ j \in\lbrace 1,..,|k_{i}|\rbrace $ , and $|k_{i}|$ is the total number of trips for site $i$.
Therefore, under the assumption of maximum depot productivity, an effective algorithm to find a feasible sequence of trips $K$ can be designed based on the following principle: given a number of trucks $m$ sufficient for the maximum depot productivity such that its upper bound $m_{u}\leq 2.\gamma.L^{-1}_{t}$, then the objective function to minimize the sites waiting time with no truck queue at sites can be stated as follows:
\\
\begin{equation}
 min \sum_{i=1}^{n}\sum_{j=1}^{|k_{i}|} k^{s}_{ij+1} - (k^{s}_{ij}+U_{i})   
\end{equation} 
\\
A feasible sequence of trips $K$ that minimizes the objective function in (18) and guarantees no trucks queue can exist in the sites can be generated as follows:
\\
\begin{equation}
 k^{s}_{ij+1}= k^{s}_{ij}+ \beta U_{i}     
\end{equation} 
\\
where $\beta$ is an optimization variable such that $\beta\geq 1$, and used to maintain the solution optimality. Therefore, the best sequence of trips is the one that is resulted by a $\beta$ satisfies (17) and (18). 

\paragraph*{ \textbf{Principle of Design}}

The principle  of design for the heuristic algorithm for the objective in (18) can be stated as follows: If two sites $i\in\lbrace a,b\rbrace$ have two trips $k_{ax}$ and $k_{by}$ where $j\in\lbrace x,y\rbrace$ and have the same loading time at the depot $ k^{d}_{ax}=k^{d}_{by}$  such that $a\neq b$, then the priority is given to the site $b$ if the site unloading time $U_{b}<U_{a}$. 

The priority can also be given according to other considerations, for example, if the a site has a specific requirement for the maximum time lagging $\gamma_{i}$ between any consecutive trips for the site  such that  $\left( k^{s}_{ij+1}- k^{s}_{ij}\right)\leq \gamma_{i}$, then the priority is given in this case for the site with the minimum  $\gamma_{i}$. Even though such a requirement is adopted by the literature, but actually in real life it is hard to claim the importance of such a requirement because usually all the sites prefer a minimum time lagging in their deliveries.

this $principle$ constitutes  logical approach to locate feasible regions in the problem solution space. Moreover, when we try to solve a dynamic problem such as RMCDP, one feasible approach for that is by giving our solution algorithm the capability to take the proper decision during the processing time. Therefore, this design principle is the criteria of such a decision. Algorithm 2 represents the $principle$ used for minimizing the sites waiting for their deliveries. 

%******************
\begin{algorithm}
\DontPrintSemicolon
\SetKwInOut{Input}{Input}\SetKwInOut{Output}{Output}
\SetKwFunction{KwFn}{}
\Input{$n,i\in\lbrace 1,..,n\rbrace,k_{i},k^{t}_{i},k^{s}_{i},D^{s},\gamma,L_{t},U_{i},m_{u} $}
\Output{$K$}
\BlankLine
%\nonl\textbf{RMCDP ($G_{|V|},H_{G}$)}\;
$K\leftarrow\phi$\;
$minWait \leftarrow\infty$\;
$Objective$ $variable $ $obj\in\left\lbrace U_{i}  \right\rbrace  $\;
$ m_{u}\leftarrow 2.\gamma.L^{-1}_{t}$\;  
$Sort \left(U_{i} \right)$ $in$ $ascending$ $order:L_{t}+\frac{d_{i}}{v_{i}}+U_{i}\leq\gamma $\; 
$Generate$ $a$ $set$ $of$ $permutations$ $P_{n}$ for $n$:
$|P_{n}|$=$\left( n\right) !$\;
\For {$each $ $p \in P_{n}$}{
{$isFeasible\leftarrow True$}\;
$W_{p} \leftarrow 0$\;
\For{$i = 1$ \KwTo $n$}{
$w_{i} \leftarrow 0$\;
$k^{d}_{i1}=D^{s}+[index\left( U_{i}\right)-1].L_{t}$\;
$k^{s}_{i1}=k^{d}_{i1}+ L_{t} + \frac{d_{i}}{v_{i}}$\;
$w^{s}_{i} \leftarrow(k^{s}_{i}<k^{s}_{i1}?k^{s}_{i1}-k^{s}_{i}:0)$\;
$ obj\leftarrow \beta U_{i} $ \;

\For {$j = 1$ \KwTo $|k_{i}|-1$}{
$k^{d}_{ij+1}= k^{d}_{ij}+ obj $\;
\If{ $IsNotAvaliable$ $\left( k^{d}_{ij+1}\right) $}
{
$k^{d}_{ij+1}\leftarrow NextEmptyLoadingTime\left( \right) $\;
\If{ $k^{s}_{ij+1}-k^{s}_{ij}\leq\gamma $}
{
$w_{i}\leftarrow w_{i}+k^{s}_{ij+1}-\left( k^{s}_{ij}+ obj\right) $\;
}
\Else{$isFeasible\leftarrow False$\; $break$}

}
}
\If{$isFeasible$}{
$W_{p}\leftarrow W_{p}+w_{i}+w^{s}_{i}$\;
}
\Else{$W_{p}\leftarrow \infty$\;$break$}
}
\If{$minWait> W_{p}$}{$minWait\leftarrow W_{p}$\;
$K\leftarrow allTripSeq(p)$
}
}

\Return $K$

\caption{RMCDP Priority Algorithm\label{IR}}
\end{algorithm}
%**************
In algorithm-2,  $w^{s}_{i}$ is site $i$ waiting time for first delivery, $w_{i}$ is site $i$ total waiting time for next deliveries, $W_{p}$ is the summation of the total sites waiting time by the possible solution $p$. The other parameters are  defined in the prior sections. The algorithm returns the best sequence of trips $K$ that has the minimum $W_{p}$ based on the objective variable $obj$ that is in use. The objective function of the problem is to minimize the sites waiting time while maintaining zero truck idling time at the sites. Therefore, the truck unloading parameter $U_{i}$ is the objective variable such that $obj=\beta U_{i}$, where $\beta$ according to (19) is an optimization factor with a value of $\beta=1$  as a default value. This default value guarantees no idling time for the trucks at the sites, and $W_{p}$ is used here for the best sequence of trips that makes the sites have minimum time when awaiting for their next trips.  The constraint in line 5 is to guarantee the accessibility for each  site $i$ from the depot with the time constraint $\gamma$. 

The advantage of the Priority Algorithm is best evidence from the huge reduction in the problem solution space $K_{s}$ to a computational cost of $ \left(n! \right)   $ where $n$ is the total number of sites to be serviced. In order to illustrate this significant reduction, let us reconsider the average problem size of 5 sites and 5 trips per site, $K_{s}$ for this realistic example is around 6 trillion possible solutions, while the  priority algorithm in this case is able to design $ \left( 5!\right) = 120$ possible solutions  100$\%$ of them are feasible and competitive as is shown in the implementation section thanks to the principles of design.
%***************
\\
\begin{table}[t]
\caption{List of variables and parameters.  } % title of Table
\centering % used for centering table

\begin{tabular}{c c l c   } % centered columns (2 columns)
\hline\hline \\ %inserts double horizontal lines
Symbol  && Description  \\[0.5ex] % inserts table
%heading
\hline \\ % inserts single horizontal line
$t$ &&Depot loading time slot \\
$D^{s}$ &&Depot starting time\\
$T_{k}$ &&All available time slots for product loading  \\  
$h_{i}$ &&Site $i$ travel time \\ 
$k^{s}_{ij}$ &&Site $i$ Trip $j$ starting time at site \\
$k^{s}_{i}$ &&Site $i$ proposed time by customer for first trip  \\
$k^{s}_{i1}$ &&Site $i$ first trip time at the site \\
$k^{d}_{ij}$ &&Site $i$ Trip $j$ starting time at depot \\
$W_{ijj+1}$ &&Site $i$ wait time between consecutive trips $j$ and $j+1$ \\
$T_{ijj+1}$ &&Site $i$ time between consecutive trips $j$ and $j+1$ \\
$W_{i}$ &&Site $i$ wait of first trip \\
$|k_{i}|$ &&Site $i$ total number of trips\\
$X_{tij}$ &&Binary variable for time slot $t$ site $i$ trip $j$\\
\\    
\hline %inserts single line 
\end{tabular}
\begin{flushleft}
\begin{center}

\end{center}

\end{flushleft}
\label{table:nonlin} % is used to refer this table in the text
\vspace{-0.35cm}

\end{table}
%-------------------------------- 
%****************************************
 \subsubsection{\textbf{Integer Programming Approach - Min Site Waiting}}
   The objective of minimizing the site waiting time for concrete delivery can be given as follows:
   \\\\
 \begin{equation}
 min \sum_{i=1}^{n}\sum_{j=1}^{|k_{i}|-1}k^{s}_{ij+1} - (k^{s}_{ij}+U_{i}) + \sum_{i=1}^{n}( k^{s}_{i1}-k^{s}_{i} )
\end{equation} 
\\
where $k^{t}_{i}$ is the trip duration of site $i$, $L_{t}$ is the loading time at the depot, $U_{i}$ is product unloading time at the site $i$, and  $|k_{i}|$ is the number of trips needed to satisfy the demand of site $i$. The proposed objective function in (22) can achieve more efficient operational management that has less overhead cost for minimizing the sites total idling time cost. 

In this section we introduce our model to solve the previous objective function. Our model is formulated as a Mixed Integer Programming (MIP) problem as follows:
\\\\
\begin{equation}
 min \sum_{i=1}^{n}\sum_{j=1}^{|k_{i}|-1}W_{ijj+1}   + \sum_{i=1}^{n}W_{i}
\end{equation} 
 \textbf{s.t.}
 \\
 \begin{equation}
k^{s}_{ij+1}-k^{s}_{ij}-T_{ijj+1}=0 \quad\quad \forall i\in \left\lbrace 1,..,n\right\rbrace 
\end{equation} 
\begin{equation}
 T_{ijj+1}-U_{i}- W_{ijj+1}=0  \quad\quad \forall j\in \left\lbrace 1,..,|k_{i}|-1\right\rbrace 
\end{equation} 
\begin{equation}
 T_{ijj+1}-U_{i}\geq0 
\end{equation} 
\begin{equation}
 T_{ijj+1}-\gamma\leq0 
\end{equation} 
 \begin{equation}
k^{s}_{i1}-k^{s}_{i}-W_{i}=0
\end{equation} 
 \begin{equation}
k^{s}_{ij}-k^{d}_{ij}-h_{i}-L_{t}=0
\end{equation} 
 \begin{equation}
\sum_{i=1}^{n}\sum_{j=1}^{|k_{i}|}\sum_{t=1}^{T_{k}}( D_{s}   + (t-1)L_{t})X_{tij} - k^{d}_{ij}=0
\end{equation} 
 \begin{equation}
\sum_{t=1}^{T_{k}} \sum_{i=1}^{n}\sum_{j=1}^{|k_{i}|}X_{tij}\leq 1
\end{equation} 
 \begin{equation}
\sum_{i=1}^{n}\sum_{j=1}^{|k_{i}|} \sum_{t=1}^{T_{k}} X_{tij}\leq 1
\end{equation} 
\begin{equation}
k^{s}_{ij},U_{i},k^{s}_{i},h_{i},L_{t}\geq 0
\end{equation} 
\begin{equation}
 X_{tij}\in \lbrace 0,1\rbrace
\end{equation}  
Definitions of the above notations are given in Table-II. The  above equations describe the model main functionality to minimize the total sites waiting time with no truck queues at sites as follows:\\

\begin{itemize} 
\itemsep0em 
 
 \item[$\bullet$ ] In (21) the objective function minimizes the sites waiting time for the first delivery, and also the time between the next consecutive trips per site.
  \item[$\bullet$ ]  In (22) the variable $T$ is defined as a time difference between the consecutive trips per site.
  \item[$\bullet$ ] In (23) the objective function variable $W_{ijj+1}$ is defined after excluding the site unloading time from the time duration between the consecutive trips for that site. 
  \item[$\bullet$ ] In (24) a constraint is enforced between the consecutive trips per site to be greater than or equal to the unloading time of that site. \emph{This constraint is to avoid the truck idling time at sites.  }
  \item[$\bullet$ ]  In (25) a constraint is enforced between the consecutive trips per site to be less than or equal to the concrete setting time. This constraint is to avoid the cold joint problem of the concrete.
  \item[$\bullet$ ] In (26) the variable $W_{i}$ is defined as a waiting time   between the first trips per site and the site proposed service starting time. 
  \item[$\bullet$ ]In (27) the variable $k^d_{ij}$ is defined as the service starting time at the depot for loading trip $j$ of site $i$.
  \item[$\bullet$ ] In (28) a binary decision variable $X_{tij}$ to assign the proper loading time slot that best fits with the problem constraints. 
  \item[$\bullet$ ] constraints (29) and (30) are to ensure that each site trip will be serviced exactly one time by exactly one loading time slot from the depot. 
 
\end{itemize} 

\section{IMPLEMENTATION AND RESULTS}
We implemented our Priority Algorithms in C/C++, using Intel 2.3 GHz CPU, 6 GB RAM, and Windows 7. The MIP model is formulated and solved using  CPLEX solver as an  optimization tool. 
Because of the absence of any standard testing datasets in the RMCDP domain, we propose different problem instances for the evaluation purpose. The instances are used to evaluate our approaches to solve the problem for the objective function of  Minimizing sites' waiting for delivery.

The scenarios beyond the proposed instances are inspired from real operational datasets, for example, the values of average instance of five sites are based on the operational data collected by the relevant professional association [1].
 We quantized the values of the instances main parameters to certain levels  in order to exclude any noisy data. In order to simplify our assumptions, we assume that the difference between the hauling and return distances can practically be neglected. 
 
  In real life, the problem is actually open to high diversity of quantities for its main parameters. Therefore, the average case scenario is the focus of this study.
\vspace{-0.1cm}  
\subsection{\textbf{Minimizing Sites Waiting}} 
Our objective function is to minimize the site waiting time for their deliveries while maintaining zero truck idling time at the sites. Each of the MIP approach and Priority algorithm are evaluated against this objective. 
The instances  are shown in Table-III and representing average and large instances respectively.

In real life, there is a range of acceptable and applicable values that can be used with the main parameters in the table for the evaluation purpose.
 % *******************************
\begin{table}[b]
\caption{The table represents the experiment parameters for three problem instances.  } % title of Table
%\vspace{-0.3cm}
\centering % used for centering table

\begin{tabular}{c c  c c } % centered columns (2 columns)
\\
\hline\hline\\   %inserts double horizontal lines
Parameter  & Instance-1   & Instance-2 & Unit  \\[0.5ex] % inserts table
%heading
\hline \\ % inserts single horizontal line
No. of Sites & 5   & 9& -\\ % inserting body of the table
Site Demand & $50^{*}$  & 50 & $m^{3}$ \\ 
RMC Setting Time & 90  &90 &$min$  \\ 
Truck Capacity & 10  & 10& $m^{3}$ \\ 
Depot Productivity & 120  &120 &$m^{3}$/h  \\ 
Truck Avg. Speed & 60  &60 &$Km/h$ \\ 
Site Starting time  & 8:00  &8:00 &$ AM$ \\
Site Unload time & 25,25,25    & 20  &$min$ \\ 
     -      &30,30   & - &-\\ 
Site Distance & 30,20,20   &30,30,30,20, &$Km$ \\ 
     -      &10,10   &20,20,10,10,10 &-\\ 
 [1ex] % [1ex] adds vertical space
  
\hline %inserts single line 
\end{tabular}
\begin{flushleft}
 
* The quantity represents each site demand.\\
%\vspace{-0.6cm}  

\end{flushleft}
\label{table:nonlin} % is used to refer this table in the text
\end{table}
 
%*********************************************
 %***************************************
%***********
\begin{figure*}[!t]
\centering
\subfloat[]{\includegraphics[scale=0.5]{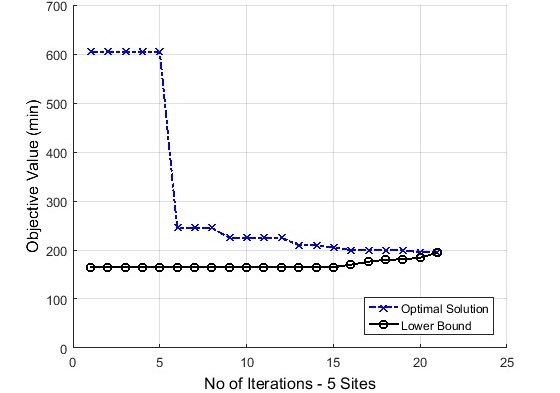}%
\label{fig_first_case0}}
\hfil
\subfloat[]{\includegraphics[scale=0.5]{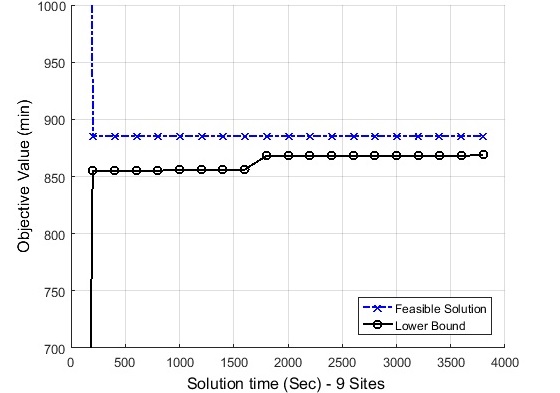}%
\label{fig_second_case2}}
%\subfloat[Case III]{\includegraphics[scale=0.2]{Per}%
%\label{fig_third_case}}
 
\caption{
CPLEX Solutions convergence for the problem instance-I and II.  }
\label{fig_sim}
%\vspace{-0.3cm}
\end{figure*}
%***************************************
%******************************* 
In Table-III, two instances are proposed for average case and large case scenarios. The parameters value are practically acceptable and matching some real life operational data that is inspired from [1], and was reported also in firm x, in Alexandria city. For both instances, We assume that the depot starting time is at 8:00 AM, and the number of trucks in use  are the upper bound according to Eq.(6).

We give a special consideration for these two possible instances. Both instances represent important scenarios show the advantage of  priority algorithm over MIP approach. In instance-1 of 5 sites, besides that it is the average case scenario in real life, it also represents the case that for any site $i$,  the  unloading time $u_{i}\geq n.L_{t}$ $ \forall i \in \left\lbrace   1,..,n  \right\rbrace $. In this case the probability of fair distribution of the loading time slots among the sites can increase. This gives the algorithm the advantage to trap the optimal solution as is shown in Table-IV.

%******************************
 \begin{table}[ht]
\caption{The table represents the optimal found by the optimizers for the problem instance-1. } % title of Table
%\vspace{-0.3cm}
\centering % used for centering table

\begin{tabular}{c c c c c c c c c c } % centered columns (2 columns)
\\
\hline\hline \\ %inserts double horizontal lines
Site-Trip & &Start at & &Start at& & End at & & Delivery \\ 
         & &  Depot   && Site    && Site &  &
\\[0.5ex] % inserts table
%heading
\hline \\ % inserts single horizontal line
1-1 && 8:00 && 8:35 && 9:00  &&10 \\
1-2 && 8:25 && 9:00 && 9:25  && 20\\
1-3 && 8:50 && 9:25 && 9:50 && 30 \\
1-4 && 9:15 && 9:50 && 10:15 && 40 \\
1-5 && 9:40 && 10:15&& 10:40 && 50 \\\\

2-1 && 8:05 &&8:30 && 8:55 && 10\\ 
2-2 && 8:30 &&8:55  && 9:20 && 20\\
2-3 && 8:55 &&9:20  && 9:45&& 30\\ 
2-4 && 9:20 &&9:45 && 10:10 && 40\\
2-5 && 9:45 &&10:10 && 10:35 && 50\\\\  

3-1 && 8:10 &&8:35 && 9:00   && 10\\ 
3-2 && 8:35 &&9:00  && 9:25  && 20\\
3-3 && 9:00 &&9:25 && 9:50   && 30\\
3-4 && 9:25 &&9:50 && 10:15  && 40\\
3-5 && 9:50 &&10:15 && 10:40 && 50\\\\

4-1 && 8:20 &&8:35 && 9:05  && 10\\ 
\textbf{4-2} && \textbf{9:05} &&\textbf{9:20}  && \textbf{9:50} && \textbf{20}\\
4-3 && 9:35 &&9:50 && 10:20&& 30\\
4-4 && 10:05 &&10:20 && 10:50&& 40\\
4-5 && 10:35 &&10:50 && 11:20&& 50\\\\

5-1 && 8:15 &&8:30 && 9:00  && 10\\ 
5-2 && 8:45 &&9:00  && 9:30 && 20\\
\textbf{5-3} && \textbf{9:30} &&\textbf{9:45} && \textbf{10:15}&& \textbf{30}\\
5-4 && 10:00 &&10:15 && 10:45&& 40\\
5-5 && 10:30 &&10:45 && 11:15&& 50\\
 [1ex] % [1ex] adds vertical space
\hline %inserts single line
\end{tabular}
\label{table:nonlin} % is used to refer this table in the text
%\vspace{-0.3cm}
\end{table}
%*******************************

For the big data obtained by solving the problem instances 2, we illustrate only the full results of problem Instance-1. Basically, Instance-1 has a total solution space of more than 600 trillions  of possible permutations according to (7). Each permutation represents a unique sequence of trips that  can be feasible or infeasible. Finding the optimal solution by trying each permutation may take at least 72 hours with our current 2.3 GHz CPU. Therefore, our MIP model and priority algorithm are designed and used as low computational cost approaches to solve the problem.

 The optimal sequence of trips for instance-1 found 
by both of CPLEX optimizer and the priority algorithm is ( 1 2 3 5 4 1 2 3 5 1 2 3 4 1 2 3 5 4 1 2 3 5 4 5 4 ). The detailed scheduling of the solution is given in Table-IV. The solution satisfies the minimum site waiting time during the product delivery and also minimum truck idling time at the sites. The optimal solution found is 195 $min$ as a minimum sites waiting time for their deliveries and zero truck waiting time at the sites.  It can be recognized from the above scheduling plan that the next trip for each site starts upon the completion of the previous trip unloading phase  which is a result of (19) in the algorithm  and  constraint (26) in our MIP model.   Also because the depot service  starting time is at 8:00 AM which is the 
same time that each site expects its first delivery according to the problem parameters in Table-III, therefore, a delay occurs for the first deliveries of the sites such that 
site-1,2,3,4 and 5  wait 35,30,35,35 and 30 min respectively before receiving their first deliveries. The other sites waiting time occur in site-4 and 5. In site-4 the delay occurs in trip-2 which starts  at the site 15 $min$ late after its previous trip ended at the site. It is important to mention again that the optimal solution shown in Table-IV is the same solution found by each of CPLEX and the priority algorithm separately.

%****************** Fig 5 ***********
In the previous section we showed that the principle of design for the case of sites' waiting gives  priority for the site of shorter unloading time if two sites trip try to start the product loading at the same available time slot. Based on this principle, the priority algorithm found that site-4 has unloading time $U_{4}$ of 30 $min$ and its trip-2 try to start at depot at 8:50 AM which is 30 $min$ after its previous trip starting time. However, site-1 trip-3 is also trying to start at 8:50 AM, and because of the priority is given to the site of shorter unloading time which is site-1 ($U_{1}=25$), site-4 trip-2 is shifted to next available loading time slot which is 8:55 AM. The waiting occurs again at 8:55 because site-2 trip-3 is trying loading at the same time and has a higher priority because $U_{2} < U_{4}$ which results in site-4 trip-2 to be  shifted again to next available slot. The process repeats again with site-3 trip-3 at 9:00 which results in site-4 trip-2  to be waiting 15 $min$ till 9:05 before it starts loading at the depot. 
%******************** 
 
\begin{table*}[ht]
\caption{The table represents optimal and feasible solutions found by CPLEX for the problem instance-1 and 2. } % title of Table
%\vspace{-0.3cm}
\centering % used for centering table

\begin{tabular}{c c c c c c c c c c c c c c c c c } % centered columns (2 columns)
\\
\hline\hline \\ %inserts double horizontal lines
\# Sites& &Total Trips & & \# Integer Var. & &  \# Constraints& &Comp. Time & & &  Bound &&& Status & & Opt Gap  \\ 
  & &  & &  & &   & & (Sec)& &Lower &   & Upper&&  & &  \% 
\\[0.5ex] % inserts table
%heading
\hline \\ % inserts single horizontal line
5 && 25 && 7295 && 448 &&3.06 &&195&   & 195&& Optimal  && 0.00\\
9 && 45 && 13131 && 576 && 3600$^{*}$    &&869&  & 885&&Feasible && 1.81 \\

  % [1ex] adds vertical space
\hline %inserts single line
\end{tabular}
\begin{flushleft}
 
* Solution convergence terminated after 6 hours.\\
%\vspace{-0.6cm}  

\end{flushleft}
\label{table:nonlin} % is used to refer this table in the text
\end{table*} 

%*************************************

Another delay also happens to site-5 trip-3 for 15 $min$ after shifting its starting time from 9:15 to 9:30. Therefore, the total sites waiting time is 35+30+35+35+30+15+15=195. This is the minimum objective value found by both CPLEX and priority algorithm. Therefore, we may claim at this point that priority algorithm was successfully able to trap the optimal solution in instance-1 huge solution space thanks to its principle of design.  This optimal solution is found by CPLEX after 21 iterations and a  CPU time of 3.06 sec as shown in Fig.5-(a). The same optimal solution found by priority algorithm after 0.104 sec which is very promised and a competitive result. The statistics of both approaches are given in Table-V and VI.  
%***************************** 

   \begin{table}[b]
\caption{The table shows the optimality of the Priority Algorithm and low computational cost.} % title of Table
\vspace{-0.3cm}
\centering % used for centering table

\begin{tabular}{c c c c c } % centered columns (2 columns)
\\
\hline\hline \\ %inserts double horizontal lines
No of  & Total Solutions & Feasibility  &Best Solution &RunTime  \\ 
 Sites &  Created by Alg. &\% & (Min)&(Sec.)   \\[0.5ex] % inserts table
%heading
\hline \\ % inserts single horizontal line
 
 5& 120& 100& 195 & 0.104 \\
 9& 362880 & 16.57 & 885&16.35 \\
 [1ex] % [1ex] adds vertical space
\hline %inserts single line
\end{tabular}
\label{table:nonlin} % is used to refer this table in the text
\end{table}
%***************************************    
Instance-2 of 9 sites represents another important scenario in which  all sites can be considered to have the same unloading time duration. In contrary to instance-1 scenario where  $u_{i}\geq n.L_{t}$, in instance-2 $u_{i}\leq n.L_{t}$. Priority algorithm was able to trap again the optimal solution  after around 16 sec, while CPLEX has a very slow convergence and could not converge to zero optimality gap after 6 hours of running time as shown in Fig. 5(b). The gap is the deference between the upper and lower bounds. When the gap is zero this means an optimal solution is found. For the priority algorithm, the optimal Sequence of Trips  is  ( 1 2 3 4 1 2 3 4 1 2 3 4 1 2 3 4 1 2 3 4 5 6 7 8 5 6 7 8 5 6 7 8 5 6 7 8 5 6 7 8 9 9 9 9 9 ). This optimal solution is found after evaluating 362880 possible permutations generated by the algorithm. 60160 permutations out of 362880 are found as feasible with a percentage of 16.57\% of the created permutations. The best feasible permutation achieves 885 min sites' waiting time which is the optimal solution.

The superior performance of the priority algorithm stems from its ability to allocate feasible regions effectively in the solution space. The algorithm does not search for these regions but actually it creates a number of permutation based on its principle of design and allocate the best among them.  
For example, the total solution space for the problem instance-1 is $6.2336074$ x $10^{14}$ possible solutions, while priority  algorithm is able to create exactly 120 solutions based on the priority principle. All of the created sequences of trips are feasible and one of them was the optimal solution as shown in Table-IV. In instance-2 of 9 sites, the total solution space of the problem is $2.3183588$ x $10^{37}$ possible solutions which is a very huge space that may take around $2.8$ x $10^{24}$ hours to evaluate all the permutations in that space. The priority algorithm is able to create 362880 which constitutes around $1.5652452$ x $10^{-30}$\% of the solution space.  16.57\% of the created permutations are feasible and one of them is the optimal solution.

\paragraph*{\textbf{Algorithm Performance Analysis}}
%*******************
\begin{figure}[b]
\centering 
  \includegraphics[scale=0.40]{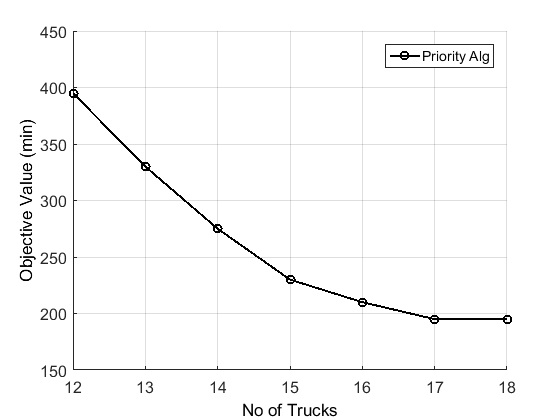}
  \caption{The impact of decreasing the number of trucks for instance-1. }
  \label{fig_inst1}

\end{figure}
%***********************
In order to analyze the performance of the priority algorithm, we evaluate its performance and optimality against range of values for instance-1 parameters. Instance-1 is in focus because it represents the average case in real life. The upper bound of the number of trucks can be used with instance-1 is 18 according to (6). Fig. 6 shows the impact of decreasing the number of trucks on the objective value.  The plot shows that decreasing the number of trucks yields an increase in the objective value. The optimal value of 195 needs 17 trucks or greater.
%***********
\begin{figure*}[!t]
\centering
\subfloat[]{\includegraphics[scale=0.31]{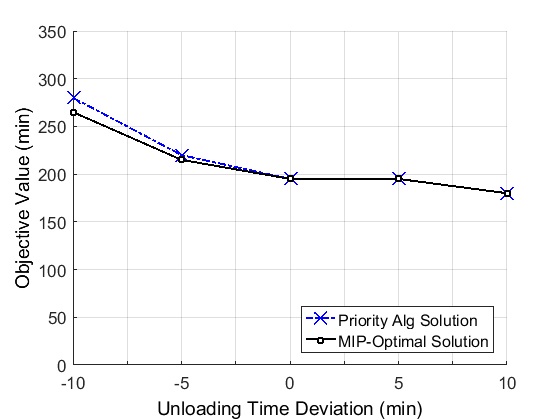}%
\label{fig_first_case0}}
\subfloat[]{\includegraphics[scale=0.31]{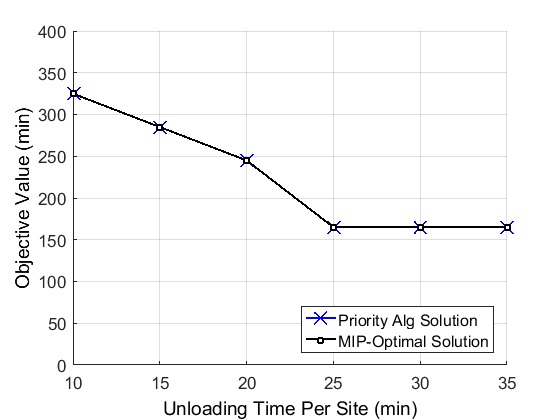}%
\label{fig_second_case2}} 
\subfloat[]{\includegraphics[scale=0.31]{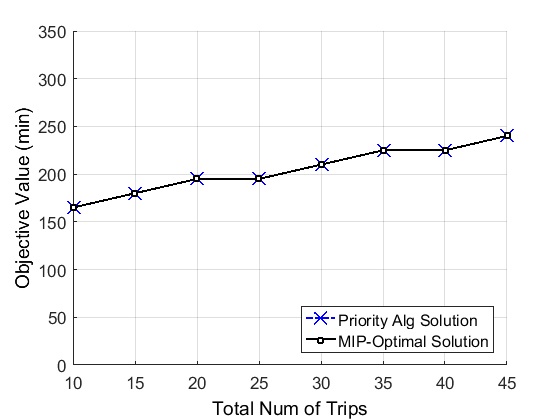}%
\label{fig_third_case2}}
\subfloat[]{\includegraphics[scale=0.31]{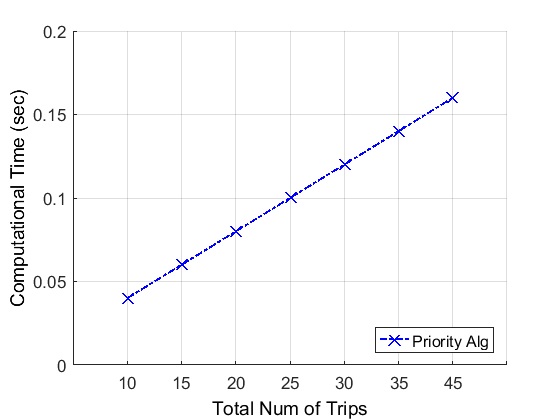}%
\label{fig_time_case1}}
\vspace{-0.3cm}
\subfloat[]{\includegraphics[scale=0.31]{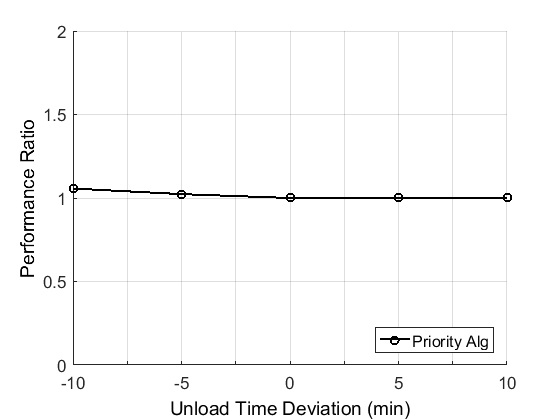}%
\label{fig_first_case12}}
\subfloat[]{\includegraphics[scale=0.31]{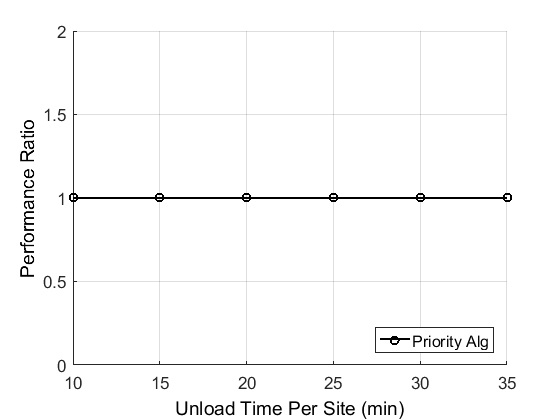}%
\label{fig_second_case22}}
\subfloat[]{\includegraphics[scale=0.31]{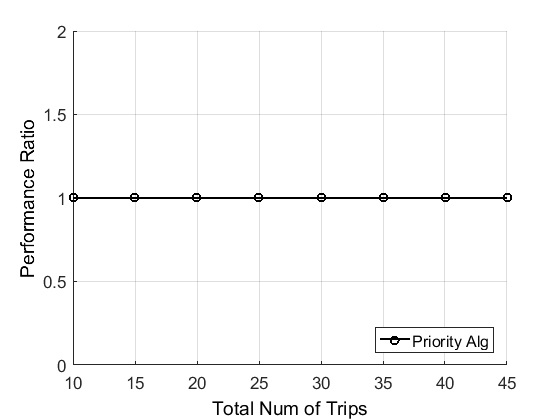}%
\label{fig_third_case32}}
\subfloat[]{\includegraphics[scale=0.31]{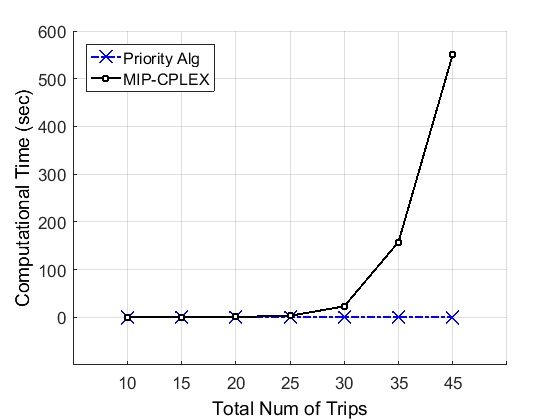}%
\label{fig_time_case2}}
%\subfloat[Case III]{\includegraphics[scale=0.2]{Per}%
%\label{fig_third_case}}
 
\caption{
Priority algorithm performance analysis for instance-I.  }
\label{fig_analysis}
\vspace{-0.3cm}
\end{figure*}
%***************************************

Because we address a minimization problem, the performance ratio $ R_{p}$ or the optimality of the priority algorithm can be defined as the ratio of the minimum objective value found by the algorithm to the optimal value such that $ R_{p}\geq 1$ . Therefore, the  minimum and best performance ratio that can be found is 1. 

Fig. 7, shows the performance analysis of the priority algorithm for instance-I. The focus on instance-I is not only because it represents the average case scenario in real life, also because it can be solved optimally in CPLEX in reasonable time. In this analysis we need to compare for each case the solution found by priority algorithm against the optimal solution found by our MIP model using CPLEX. 

 In Fig.7(a), we show the impact of increasing and decreasing the sites unloading time in table-III. We refer to these changes as deviations from the original values in the table. For example, 0 deviation in the figure means no change in the parameter values, while -5 means decreasing the unloading time for each site by 5 such that the new parameter values become $\lbrace 20,20,20,25,25\rbrace$.  The deviation of 0 here represents the original values of $\lbrace 25,25,25,30,30\rbrace$ which is the case that for any site $i$  the  unloading time $u_{i}\geq n.L_{t}$ $ \forall i \in \left\lbrace   1,..,n  \right\rbrace $. The deviation of -5 and -10 represent the cases when there is at least $u_{i} < n.L_{t}$. For example  with -5 deviation, site 1,2, and 3 each has unloading time of 20 while $n.L_{t}$=5*5=25.   For the cases of  0, 5, and 10 deviations in which $u_{i}\geq n.L_{t}$, priority algorithm is able to find the same optimal solution as CPLEX. The  results of 0 deviation is shown in Table-IV and a detailed explanation of  the algorithm  mechanism is given. The algorithm performance ratio for these deviations is given in Fig.7(e). A performance ratio of 1 is achieved for the cases $u_{i} \geq n.L_{t}$, while for cases of $u_{i} < n.L_{t}$ as in -10 and -5 deviations, performance ratios of 1.056 and 1.02 are achieved for each respectively. 

Another scenario that also common in real life is when the sites have unloading times close to each other such that the differences are slight and can be neglected.  We analyze also this case with a range of unloading times as shown if Fig.7(b) and (f). The x-axis for both graphs represents  the
unloading time per site. For example, if this time is 25 min, this means the unloading time for each of the five sites is 25 min. In this scenario, the priority algorithm is able to find optimal solutions for all unloading time values and achieves a performance ratio of 1 as shown in (f).  This algorithm advantage is best deployed in case of large instances such as instance-II of nine sites.  In instance-II the algorithm is able to find the optimal solution in 16.35 sec as shown in table-VI, while CPLEX run 3600 sec to find a feasible solution and may need days before it can  converge to optimal solution.

The computational cost of the priority algorithm is also evaluated for instance-I by changing the quantity of demands of the sites. The increase in these demands yields increasing in the total number of trips up to 45 trips as shown in Fig.7(c). By changing only, the total number of trips and keeping other parameters the same, priority algorithm is able to find optimal solutions for any number of trips as long as for any site $i$ the unloading time $u_{i}\geq n.L_{t}$. Under this condition, priority algorithm achieves a performance ratio of 1 as shown in Fig.7(g). Even though both of priority algorithm and MIP model have the same solutions as in Fig.7(c), they have completely different computational times as shown in Fig.7(h). The computational time of priority algorithm  increases almost linearly in (d) when the total number of trips increases and has very low runtime. For CPLEX, its computational time increases exponentially as shown in (h). For the case of 45 trips, priority algorithm needs around 0.16 sec to find the optimal solution, while CPLEX needs 550.9 sec to find the optimal solution.

Both of priority approach, and MIP approach are trying to allocate the feasible regions in the solution space.  The main advantages of the priority algorithm  is the optimality and low runtime it has. The main difficulty is that we need to perform deep analysis of the problem prior to designing the 
algorithm.   For the LP based solutions, they are highly time and resource consuming approaches with no guarantee of optimality when the instances become larger in size. Unfortunately, the tendency in the RMC  industry [18] indicates that its instances are going to be larger with time due to the increasing rate of the CBP productivity which  doesn't meet the same rate in the truck capacity, therefore, 
heuristic approaches such as priority algorithm are  expected to dominate in the near future and their parallel versions [19-20] may play the key roles.

\section{Conclusion}
  In this paper, the vehicle scheduling problem under capacity and time window constraints has been proposed and analyzed in depth. We adopt the RMCDP as case of study in this category and show the 
proper projection of the RMCDP in graph theory,  which is an important step that constitutes our first contribution in that domain. By this projection and proving the NP-
Completeness of the RMCDP we opened the door to import any improvement in the graph-based optimization technique to that domain and vice versa. Apart from the over complexity and high computational time of 
the linear programming based approaches, and the low optimality and feasibility of the evolutionary algorithms, we adopted the heuristic approach for its high optimality and low 
time complexity which have been proven by our results. In order to design an effective heuristic approach, we proposed the problem  set of 
definitions and  their associated analysis. Our approach is based on mining the problem main characteristics in order to design the feasible solutions in a systematic way rather than searching randomly for 
them. The algorithm shows high optimality and low runtime cost for both  average and large instances. However, we believe our approach has a full range of potential that need to be explored. Our contributions in this paper can be applied to other fields and may not be restricted to the RMCDP domain. Extending our work to other open optimization problems is under consideration as a future work.
  
%***************************************  
%[at the cost of providing solutions which in all cases is guaranteed to be optimal or only slightly sub-optimal]

%The purpose of this analysis is to show the advantages of  priority algorithm in terms of optimality and speed.
%Total solution space of instance-1 is $6.2336074x10^{14}$ possible permutation

%Priority algorithm search only $1.925049x10^{-11}$\%

%Total solution space of instance-2 is $2.3183588x10^{37}$

%Priority algorithm search only $1.5652452x10^{-30}$\%

%*************************************************  

% conference papers do not normally have an appendix

% use section* for acknowledgement
% trigger a \newpage just before the given reference
% number - used to balance the columns on the last page
% adjust value as needed - may need to be readjusted if
% the document is modified later
%\IEEEtriggeratref{8}
% The "triggered" command can be changed if desired:
%\IEEEtriggercmd{\enlargethispage{-5in}}

% references section

% can use a bibliography generated by BibTeX as a .bbl file
% BibTeX documentation can be easily obtained at:
% http://www.ctan.org/tex-archive/biblio/bibtex/contrib/doc/
% The IEEEtran BibTeX style support page is at:
% http://www.michaelshell.org/tex/ieeetran/bibtex/
%\bibliographystyle{IEEEtran}
% argument is your BibTeX string definitions and bibliography database(s)
%\bibliography{IEEEabrv,../bib/paper}
%
% <OR> manually copy in the resultant .bbl file
% set second argument of \begin to the number of references
% (used to reserve space for the reference number labels box)

% ********************** New Table ********

\end{document}